\documentclass{aa} 

\usepackage{graphicx} 
\usepackage{txfonts} 
\usepackage{natbib} 
 
\newcommand{\Mpc}{$h^{-1}$\thinspace Mpc}

\def\apj{ApJ} 
\def\apjl{ApJL} 
\def\apjs{ApJS} 
 
\def\aap{A\&A} 
\def\mnras{MNRAS}

\begin{document}    
 
\title{Extended percolation analysis of the cosmic web }
 
\author{J. Einasto\inst{1,2,3} 
\and  I. Suhhonenko\inst{1} 
\and  L. J. Liivam\"agi\inst{1}
\and   M. Einasto\inst{1}  
}
 
\institute{Tartu Observatory, 61602 T\~oravere, Estonia 
\and  
ICRANet, Piazza della Repubblica 10, 65122 Pescara, Italy 
\and 
Estonian Academy of Sciences, 10130 Tallinn, Estonia
} 
 
\date{ Received March 13, 2018; accepted May 15, 2018}  
 
\authorrunning{Einasto et al.} 
 
\titlerunning{Percolation analysis} 
 
\offprints{J. Einasto, e-mail: jaan.einasto@to.ee} 
 
\abstract {}
{We develop an extended percolation method to allow the comparison of
  geometrical properties of the real cosmic web with the simulated
  dark matter web for an ensemble of over- and under-density systems.}
{We scan density fields of dark matter (DM) model and SDSS 
  observational samples, and find connected over- and under-density
  regions in a large range of threshold densities.  Lengths, filling factors
  and numbers of largest clusters and voids as functions of the
  threshold density are used as percolation functions.  }
{We find that percolation functions of DM models of different box
  sizes are very similar to each other.  This stability suggests that
  properties of the cosmic web, as found in the present paper, can be
  applied to the cosmic web as a whole.  Percolation functions depend
  strongly on the smoothing length. At  smoothing length
  1~\Mpc\ the percolation threshold density for clusters is
  $\log P_C = 0.718 \pm 0.014$, and for voids is
  $\log P_V = -0.816 \pm 0.015$, very different from percolation
  thresholds for random samples, $\log P_0 = 0.00 \pm 0.02$.}
{The extended percolation analysis is a versatile method to study
  various geometrical properties of the cosmic web in a wide range of
  parameters.  Percolation functions of the SDSS sample are very
  different from percolation functions of DM model samples.  The SDSS
  sample has only one large percolating void which fills almost the
  whole volume.  The SDSS sample contains numerous small isolated
  clusters at low threshold densities, instead of one single
  percolating DM cluster.  These differences are due to the tenuous dark
  matter web, present in model samples, but absent in real
  observational samples.  }


\maketitle

\section{Introduction}

Studies of the 3-dimensional distribution of galaxies show that
galaxies and clusters of galaxies are not randomly clustered.  Chains
and filaments connect galaxies and clusters to a continuous
supercluster-void network \citep{Joeveer:1977, Joeveer:1978dz,
  Joeveer:1978pb, Gregory:1978, Einasto:1980, Zeldovich:1982kl},
called presently the cosmic web \citep{Bond:1996fv}.  The cosmic web
is a complex system, and there exists a rapidly growing number of
methods to describe the web in quantitative terms.  Recent reviews of
methods to characterise the structure of the web were given among
others by \citet{van-de-Weygaert:2009bh}, \citet{Aragon-Calvo:2010wd},
and \citet{Libeskind:2017fk}. An overview of the present status of the
study of the cosmic web can be find in the Zeldovich Symposium
proceedings \citep{VandeWeygaert:2016zt}.  

One essential geometrical property of the cosmic web is the
connectivity of components of the web: clusters are connected by
filaments to superclusters and to the whole web, similarly voids form
a complex system connected by tunnels.  The connectivity property is
analysed in the percolation theory, and is applied in physics,
geophysics, medicine etc.; for an introduction to the percolation
analysis see \citet{Stauffer:1979aa}.  The percolation method was
introduced in cosmological studies by \citet{Zeldovich:1982kl},
\citet{Melott:1983}, and \citet{Einasto:1984zh}. Its principal idea
was explained by \citet{Shandarin:1983kx}.  In these first studies the
percolation method was applied to particles and galaxies.

A natural extension of the method is to use instead of particles the
density field, which allows to study the connectivity of over- and
under-dense regions.  Using an appropriate threshold density the field
is divided into high-density and low-density regions. Connected
high-density regions are called clusters, and connected low-density
regions are called voids.  If the threshold is high, then clusters are
small, and are isolated from each other.  When the threshold density
decreases then clusters start to merge, and at certain threshold
density the largest cluster spans the whole volume under study. This
threshold is called percolating density threshold.  Similarly the
connectivity of voids can be investigated.  The percolation method
using density fields was applied by \citet{Einasto:1986a,
  Einasto:1987kw}, \citet{Boerner:1989aa, Mo:1990aa},
\citet{Dominik:1992eu}, \citet{Klypin:1993aa}, \citet{Yess:1996aa,
  Yess:1997aa}, \citet{Sahni:1997ai}, \citet{Sathyaprakash:1998aa,
  Sathyaprakash:1998ab}, \citet{Shandarin:2004ij},
\citet{Shandarin:2006bs}, \citet{Einasto:2014aa}. { Percolation
  processes were also used to identify elements of the cosmic web.
  \citet{Aragon-Calvo:2007} applied the Multiscale Morphology Filter
  to identify clusters, filaments and walls of the cosmic
  web.  \citet{Cautun:2013aa, Cautun:2014qy} developed the NEXUS and
  NEXUS+ algorithms to identify filaments and sheets by finding the
  threshold where filament and sheet networks percolate.}

So far the percolation analysis was concentrated to the study of
properties of clusters and voids near the percolating threshold
density. Most percolation studies were applied to the study of
connectivity of simulated dark matter (DM) samples only.  The goal of
this paper is to develop and test a modification of the percolation
method.  The extended version of the percolation analysis differs from
most previous percolation analyses as follows: (i) we use a {\em wide}
threshold density interval to find cluster/void lengths and filling
factors; (ii) we use a {\em large range} of smoothing lengths to
describe the density field of DM and galaxies in a complex way; (iii)
we apply the percolation analysis to {\em compare DM models with
  observations}.  { We use only positional data, available for DM
  particles and galaxies, and ignore velocities, not available for
  galaxies.}  We consider DM as a physical fluid having continuous
density distribution, thus simulated DM particles are only markers of
the field.  Similarly we consider observed galaxies as markers of a
smooth luminosity density field.

We divide the cosmic web under study at each threshold density into
high- and low-density systems, clusters and voids.  For each threshold
density we find catalogues of clusters and voids, and select the
largest clusters and largest voids. Lengths and volumes of largest
clusters and voids, and numbers of clusters and voids at respective
threshold density level, as functions of the threshold density, are
used as {\em percolation functions}.  Percolation functions allow an easy,
very compact and intuitive presentation of general geometrical
properties once for the whole web --- ensembles of all clusters and
voids for a particular parameter set.  Catalogs of clusters and
voids are essential parts of the method, and provide information on
individual clusters and voids.

We use the density field estimator with a constant grid, as
applied in numerical simulations of the evolution of the cosmic web.
The density field found in numerical simulations presents the ``true''
density of the DM model.  To compare models with observations
it is important to apply a proper smoothing level suited for a
particular task.  We shall use smoothing kernel sizes from 1 to 
8~\Mpc\ to see the effect of smoothing to geometrical properties of
the web.

To see the dependence of percolation properties on the size of the
sample, we shall use numerical simulations of the evolution of the web
applying $\Lambda$CDM cosmology in boxes of sizes from 100 to
1024~\Mpc.  In all models we use cosmological parameters: Hubble
parameter $H_0 = 100 h$~km~s$^{-1}$~Mpc$^{-1}$, matter density
parameter $\Omega_{\mathrm{m}} = 0.28$, and dark energy density
parameter $\Omega_{\Lambda} = 0.72$.  Density fields of simulated
$\Lambda$CDM samples are given for grid sizes from $\sim 0.2$ to
2~\Mpc. This is sufficient to investigate global geometrical
properties of our observed and simulated samples. For comparison we
use the main sample of the SDSS DR8 survey to calculate the luminosity
density field of galaxies.

The paper is organized as follows. In the next Section we describe the
calculation of the density field of observed and simulated samples,
and the method to find clusters, voids and their parameters.  In
Section 3 we perform percolation analysis of DM simulated clusters and
voids.  In Section 4 we compare percolation properties of model and
SDSS samples. The last Section brings the general discussion and
summary remarks.

\section{Data and methods}

\subsection{Simulations of the evolution of the  cosmic web}

Simulations of the evolution of the cosmic web were performed in boxes
of sizes $L_0=100,~256,~512$,~1024~\Mpc, with resolution
$N_{\mathrm{grid}} = 512$ and
$N_{\mathrm{part}} = N_{\mathrm{grid}}^3$ particles.  This wide
interval of simulation boxes was used to investigate the influence of
the simulation box to percolating properties of models.  We designate
the simulations as L100, L256, L512, and L1024.  The initial density
fluctuation spectra were generated using the COSMICS code by
\citet{Bertschinger:1995}.  We assumed cosmological parameters
$\Omega_{\mathrm{m}} = 0.28$, $\Omega_{\Lambda} = 0.72$,
$\sigma_8 = 0.84$, and the dimensionless Hubble constant $h = 0.73$.
To generate initial data we used the baryonic matter density
$\Omega_{\mathrm{b}}= 0.044$ (\citet{Tegmark:2004}).  Calculations
were performed with the GADGET-2 code by \citet{Springel:2005}.
Particle positions  were extracted for seven epochs
between redshifts $z = 30 - 0$.  The cell size of the simulation L512
is $L_0/N_{\mathrm{grid}} =1$~\Mpc, identical with the size of cells
of the density field of observational SDSS main sample of galaxies,
used by \citet{Liivamagi:2012aa}.

\subsection{SDSS data} 

The density field method allows to use flux-limited galaxy samples,
and to take statistically into account galaxies too faint to be
included to the flux-limited samples, as applied among others by
\citet{Einasto:2003a, Einasto:2007tg}, and \citet{Liivamagi:2012aa} to select
superclusters of galaxies. 

We use the Sloan Digital Sky Survey (SDSS) Data Release 8 (DR8)
\citep{Aihara:2011aa} and galaxy group catalogue by
\citet{Tempel:2012aa} to calculate the luminosity density field.  In
the calculation of the luminosity density field we need to take into
account the selection effects that are present in flux-limited samples
\citep{Tempel:2009sp, Tago:2010ij}. In the calculation of the
luminosity density field galaxies were selected within the apparent
{\em r} magnitude interval $12.5 \le m_r \le 17.77$
\citep{Liivamagi:2012aa}.  In the nearby region relatively faint
galaxies are included to the sample, in more distant regions only the
brightest galaxies are seen.  To take this into account, we calculate
a distance-dependent weight factor:
\begin{equation}
  W_L(d) =  {\frac{\int_0^\infty L\,\phi(L)\
      \mathrm{d}L}{\int_{L_1}^{L_2} L\,\phi(L)\,\mathrm{d}L}} ,
  \label{eq:weight}
\end{equation}
where $L_{1,2}=L_{\sun} 10^{0.4(M_{\sun}-M_{1,2})}$ are the luminosity
limits of the observational window at distance $d$, corresponding to
the absolute magnitude limits of the window $M_1$ and $M_2$.  The
weight factor $W_L(d)$ increases to $\approx 8$ at the far end of the
sample; for a more detailed description of the calculation of the
luminosity density field and corrections used see
\citet{Liivamagi:2012aa}.

\subsection{Calculation of the density field} 
 
In numerical simulations of the evolution of the cosmic web for each
simulation step the density field with resolution
$L_0/N_{\mathrm{grid}}$~\Mpc\ is calculated to find the gravitational
potential field and vice versa.  We extracted particle positions and
density fields at each simulation epoch, which represent true
densities of our DM models.

We determined  smoothed density fields of galaxies and simulations
using a $B_3$ spline \citep[see][]{Martinez:2002fu}:
\begin{equation} 
B_3(x)=\frac1{12}\left[|x-2|^3-4|x-1|^3+6|x|^3-4|x+1|^3+|x+2|^3\right]. 
\end{equation} 
The spline function is different from zero only in the interval
$x\in[-2,2]$.  To calculate the high-resolution density field we use
the kernel of the scale, equal to the cell size of the simulation,
$L_0/N_{\mathrm{grid}}$, where $L_0$ is the size of the
simulation box, and $N_{\mathrm{grid}}$ is the number of grid elements
in one coordinate.  The smoothing with index $i$ has a smoothing
radius $r_i= L_0/N_{\mathrm{grid}} \times 2^i$. The effective scale of
smoothing is equal to $2\times r_i$.  

To investigate the influence of the smoothing length we calculated
density fields with smoothing up to index 5.  For the L100 model
smoothing with indexes $i=2$, 3, 4, and 5 corresponds to kernels of
radii 0.78, 1.56, 3.125, 6.25~\Mpc, for the L256 model indexes $i=1$,
2, 3, 4 correspond to kernels of radii $R_B =1$, 2, 4 and 8~\Mpc, for
the L512 model smoothing indexes $i=1$, 2 and 3 correspond to 
kernels of radii $R_B =2$, 4 and 8~\Mpc, and for the L1024 model
indexes $i=1$, 2 correspond to kernels of radii 4 and 8~\Mpc.  The
comparison between $B_3$ spline kernel and Gaussian kernel is given in
Appendix C of \citet{Tempel:2014uq}. The $B_3$ kernel of radius
$R_B=1$~\Mpc\ corresponds to a Gaussian kernel with dispersion
$R_G = 0.6$~\Mpc.

\subsection{Finding clusters and voids}
 
The main step in the percolation method is finding of over-density and
under-density regions of the density field. We call over-density
regions geometrical clusters, and under-density regions geometrical
voids, or shortly clusters and voids.  The difference between
geometrical clusters and voids, and physical clusters and voids shall
be discussed below.  In cluster search we use several loops over the
density field. The first loop is over threshold densities. 

We scan the density field in a range of threshold densities from
$D_t=0.1$ to $D_t=10$ in mean density units.  For our study the
behaviour of voids is critical, thus we use a logarithmic step of
densities, $\Delta \log D_t = 0.02$, to find over- and under-density
systems. In this way there is the same number of steps in regions
below and above the mean density level.  This range covers all
densities of practical interest, since in low-density regions the
minimal density is $\approx 0.1$, and the density threshold to find
conventional superclusters is $D_t \approx 5$
\citep{Liivamagi:2012aa}.  We mark all cells with density values equal
or above the threshold $D_t$ as ``filled'' regions, and all cells
below this threshold as ``empty'' regions.

Inside the first loop we make another loop over all ``filled'' cells
to find neighbours among ``filled'' cells. Two cells of the same type
are considered as neighbours (``friends'') and members of the cluster
if they have a common sidewall. Every cell can have at most six cells
as neighbours; in percolation theory this is called site percolation
\citep{Klypin:1993aa}.  Members of clusters are selected using a
``Friend-of-Friend'' (FoF) algorithm: the ``friend'' of my ``friend''
is my ``friend''.

When a cluster is found, the next step is the calculation of its
parameters.  We calculate the following parameters: centre
coordinates, $x_c, y_c, z_c$ (mean values of extreme $x, y, z$
coordinates); sizes along coordinate axes,
$\Delta x,~ \Delta y,~ \Delta z$ (differences between extreme
$x, y, z$ coordinates); geometrical diameters,
$L_{\mathrm{geom}} = \sqrt{\left[(\Delta x)^2 + (\Delta y)^2 + (\Delta
    z)^2\right]}$; maximal sizes along coordinate axes,
$L_{\mathrm{max}} = \max(\Delta x,\Delta y,\Delta z)$; volumes, $V_C$,
defined as the volume in space where the density is equal or greater
than the threshold density $D_t$; total masses (or luminosities),
$M_t$, i.e. the masses (luminosities) inside the density contour $D_t$
of the cluster, both in mean density units.

During the cluster search we find the cluster with the largest volume
for the given threshold density. We store in a separate file for each
threshold density the number of clusters found, and data on the
largest cluster: the geometrical diameter, the maximal size along
coordinate axes, the volume, and the total mass (luminosity).
Diameters and maximal sizes are expressed in units of the sample size,
$L_0$ (the effective side-length in the case of the SDSS sample), the
volume (actually the filling factor) is expressed in units of the
volume of the whole sample, $V_0$.  Maximal sizes (lengths) of largest
clusters, $\mathcal{L}(D_t) = L_{\mathrm{max}}/L_0$, filling factors
of largest clusters, $\mathcal{F}(D_t) = V_{\mathrm{max}}/V_0$, and
numbers of clusters at the threshold density, $\mathcal{N}(D_t)$, as
functions of the threshold density, $D_t$, are percolation functions
to characterise general geometrical properties of the web.  If the
cluster spans the whole volume under study, $L_{\mathrm{max}} = L_0$,
the cluster is called percolating. The percolation threshold density,
$P = D_t$, is defined as follows: for $D_t \le P$ there exists one and
only one percolating cluster, for $D_t > P$ there are no percolating
clusters \citep{Stauffer:1979aa}.

A similar procedure is used to find voids.  A loop over all ``empty''
cells is made to find neighbours among other ``empty'' cells.  The
search for neighbours is made exactly the same way as the search of
over-density regions.  Parameters of voids are found using the same
procedure: the procedure uses as input only the catalogue of marked
cells, either over-density or under-density cells.  As in the case of
clusters we find for each threshold density the largest voids, and
store in a separate file the number of voids at this threshold, and
parameters of largest voids. Lengths and filling factors of largest
voids, and numbers of voids as functions of the threshold density are
percolation functions of voids.  The percolation threshold $P$ of
voids is defined inversely: for $D_t \ge P$ there exists one
percolating void, for $D_t < P$ there are no percolating voids.

During the search of high- and low-density systems we exclude very
small systems, to avoid the contamination of cluster and void
catalogues with very small systems.  We made for most samples
cluster/void search twice, using exclusion volume limits,
$N_{\mathrm{lim}} = 50$ and 500 computation cells. For the geometry
study we use mostly the largest system in each cluster and void
catalogue; the length function $\mathcal{L}(D_t)$ and the filling
factor function $\mathcal{F}(D_t)$ are not influenced by the choice of
$N_{\mathrm{lim}}$.  Clusters
and voids found for close threshold densities have usually rather
similar properties.  But close to the percolation threshold density of
clusters (voids), found for neighbouring $D_t$ values, have rather
different lengths and volumes; here percolation functions change
rapidly with $D_t$.

{ Our scanning procedure of density fields is constructed in a way
  that every grid cell is classified as being part of a cluster or
  void.  Most cells are classified as members of clusters at one
  threshold density value, and as members of voids at another
  threshold density value.  The procedure is different from
  conventional ones, where a cell can be part of only one type
  of structure element (node, wall, filament or void)
  \citep{Cautun:2014qy}.} In total we have 101 steps, and find for
each smoothing length 101 catalogs of clusters and 101 catalogs of
voids.  As we have four models with different size, four smoothing
lengths, and clusters and voids separately, we have 16 percolation
function pairs, and $4\times 4 \times 2 \times 101 = 3232$
cluster/void catalogues.  Catalogues of clusters and voids for each
search parameter contain large quantities of information.  These
catalogues were also stored, and selectively used in the present
paper.

\begin{figure*}[ht]
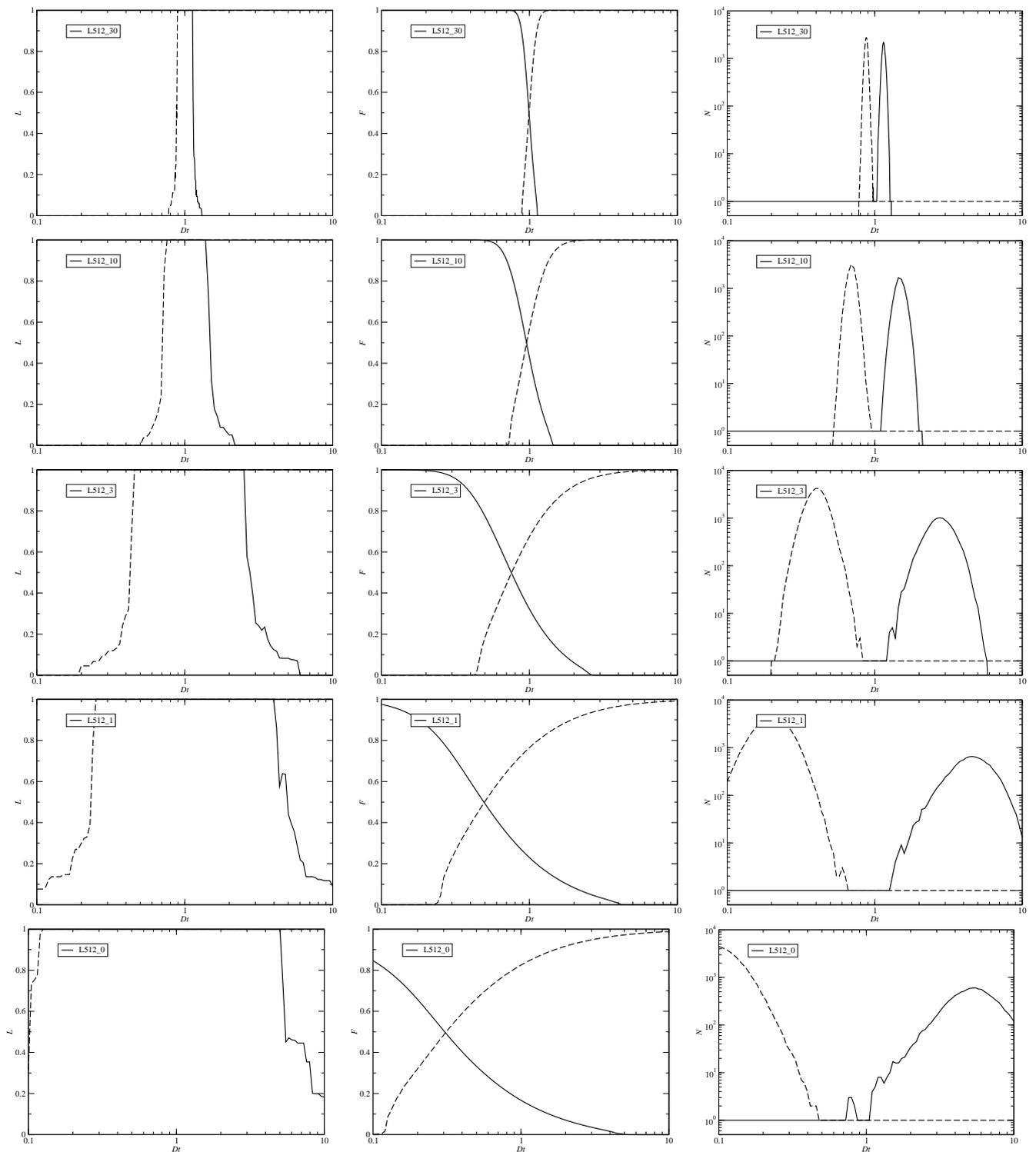
 
\centering 
\hspace{2mm}  
\resizebox{0.30\textwidth}{!}{\includegraphics*{L512_30_Lmax-D0_1.eps}}
\hspace{2mm}  
\resizebox{0.30\textwidth}{!}{\includegraphics*{{L512_30_Vmax-D0_1.eps}}}
\hspace{2mm}  
\resizebox{0.30\textwidth}{!}{\includegraphics*{L512_30_N-D0_1.eps}}\\
\hspace{2mm}  
\resizebox{0.30\textwidth}{!}{\includegraphics*{L512_10_Lmax-D0_1.eps}}
\hspace{2mm}  
  \resizebox{0.30\textwidth}{!}{\includegraphics*{{L512_10_Vmax-D0_1.eps}}}
\hspace{2mm}  
\resizebox{0.30\textwidth}{!}{\includegraphics*{L512_10_N-D0_1.eps}}\\
\hspace{2mm}  
\resizebox{0.30\textwidth}{!}{\includegraphics*{L512_3_Lmax-D0_1.eps}}
\hspace{2mm} 
\resizebox{0.30\textwidth}{!}{\includegraphics*{L512_3_Vmax-D0_1.eps}}
\hspace{2mm} 
\resizebox{0.30\textwidth}{!}{\includegraphics*{L512_3_N-D0_1.eps}}\\
 \hspace{2mm}
\resizebox{0.30\textwidth}{!}{\includegraphics*{L512_1_Lmax-D0_1.eps}}
\hspace{2mm}  
 \resizebox{0.30\textwidth}{!}{\includegraphics*{L512_1_Vmax-D0_1.eps}}
 \hspace{2mm}
  \resizebox{0.30\textwidth}{!}{\includegraphics*{L512_1_N-D0_1.eps}}
 \hspace{2mm}
\resizebox{0.30\textwidth}{!}{\includegraphics*{L512_0_Lmax-D0_1.eps}}
\hspace{2mm}  
 \resizebox{0.30\textwidth}{!}{\includegraphics*{L512_0_Vmax-D0_1.eps}}
 \hspace{2mm}
  \resizebox{0.30\textwidth}{!}{\includegraphics*{L512_0_N-D0_1.eps}}
\\ 
\caption{Change of percolation functions of the model L512 with
  simulation epoch.  Panels from top to bottom are for the initial
  epoch, $z=30$, and epochs $z=10$, $z=3$,  $z=1$, and $z=0$.  {\em Left} ---
   lengths of largest clusters and voids,
  $\mathcal{L}(D_t) = L_{\mathrm{max}}/L_0$; {\em Center} --- 
  filling factors of largest clusters and voids,
  $\mathcal{F}(D_t) = V_{\mathrm{max}}/V_0$; {\em Right}  --- numbers
  of clusters and voids, $\mathcal{N}(D_t)$; all as functions of the
  threshold density, $D_t$.  Functions are found for original
  non-smoothed density fields, which correspond to a resolution
  1~\Mpc.  Functions for clusters are plotted with solid lines, for
  voids with dashed lines.  }
\label{fig:L512evol} 
\end{figure*} 

\section{Extended percolation analysis of DM model samples}

In this Section we apply the extended percolation method to analyse
properties of DM model samples. We shall focus on three problems: How
percolation properties evolve with time? { How smoothing length
  influences geometrical properties of the density field? } and What
are principal similarities and differences of properties of the
ensemble of clusters from properties of the ensemble of voids?

\subsection{Geometrical vs physical clusters and voids} 
 
As traditional in percolation analyses high-density regions are called
clusters, and low-density regions voids.  Terms ``clusters'' and
``voids'' in our context are geometrical clusters and geometrical
voids.  Geometrical clusters can have as sub-clusters physical
clusters or superclusters, connected by filaments and sheets.  Similarly
geometrical voids may consist of physical voids, connected by
intermediate-density tunnels.  Physical clusters may be defined as
compact galaxy systems \citep{Huchra:1982fk, Tago:2010ij,
  Tempel:2014uq}, and physical voids as low-density regions,
surrounded by galaxies \citep{Einasto:1989cr, Colberg:2008fy}.
Alternatively physical clusters/superclusters can be defined by the
velocity inflow \citep{Tully:2014}, and physical voids as
single-stream regions that have not undergone shell-crossing
\citep{Falck:2015pd}, or by velocity outflow \citep{Sorce:2016ve}.  An
universal physical void identification tool VIDE was suggested by
\citet{Sutter:2015aa}. For the difference between geometrical and
physical clusters and voids see also \citet{Dominik:1992eu} and
\citet{Sheth:2003sp}.

\subsection{Percolation functions at various stages of the evolution
  of the cosmic web}
 
To understand geometrical properties of the cosmic web at various
stages of its evolution we calculated percolation functions for five
epochs of the evolution of the cosmic web, corresponding to redshifts
$z=30,~10,~3,~1$, and the present epoch, $z=0$.  Functions were
calculated for the model L512, using original density fields without
additional smoothing, as found from simulations.  During the search of
over- and under-density regions the limit to exclude very small
systems was in these calculations set to $N_{lim} =500$ computation
cells, i.e cubic \Mpc. For this reason total numbers of clusters and
voids are smaller than in most other calculations, as seen in
Table~\ref{Tab1}. Properties of largest clusters and voids are not
influenced by this difference.

{\scriptsize 
\begin{table}[ht] 
\caption{Percolation parameters of  model and SDSS clusters and voids.} 
\tiny 
\begin{tabular}{llrrrrr}
\hline  \hline
Sample&&Clusters&&&Voids \\ 
\hline  
              & $P_C$ &$N_{C}$ & $F_{C}$& $P_V$  &$N_{V}$ & $F_V$\\  
\hline  
(1)&(2)&(3)&(4)&(5)&(6)&(7)\\ 
\hline  
\\ 
L512-30  & 1.127 &  1494 &  0.0324  &   0.891   &  2111 & 0.0351\\ 
L512-10  & 1.38 & 1026 &  0.0472 &   0.759   &  1215 & 0.1282\\ 
L512-03  & 2.51 &  896 &  0.0102  &   0.457   &  2923 & 0.0690\\ 
L512-01  & 3.98 &  552 &  0.0078  &   0.251    &  3190 & 0.0478\\ 
L512-00  & 5.01 &  586 &  0.0010  &   0.126   &  2988 & 0.0864\\ 
\\ 
L100.1   & 6.61 &  1501 &  0.0050  &   0.151   &  1289 & 0.0891\\
L100.2   & 4.37 &   551 &  0.0163  &   0.182   &   537 & 0.0616\\ 
L100.4   & 3.47 &  132 &  0.0232  &   0.275   &   117 &  0.0505\\ 
L100.8   & 2.63 &   21 &  0.0386  &   0.417    &    32 & 0.0832\\ 
\\ 
L256.1   & 4.79 &  8452 &  0.0043  &   0.158   &  9207 & 0.0485\\
L256.2   & 3.98 &  4101 &  0.0065  &   0.219   &  3144 & 0.0532\\ 
L256.4   & 2.88 &  1035 &  0.0106  &   0.302   &   862 &  0.0573\\ 
L256.8   & 2.09 &   237 &  0.0253  &   0.436    &   181 & 0.0393\\ 
\\ 
L512.1   & 5.01 &  8286 &  0.0010  &   0.126   &  18399 & 0.0864\\
L512.2   & 3.80 &  10676 &  0.0037  &   0.240   &  9672 & 0.1562\\ 
L512.4   & 2.75 & 5430 &  0.0080  &   0.302   &  4464 &  0.0504\\ 
L512.8   & 2.00 &  1362 &  0.0170  &   0.398    &  1413 & 0.0127\\ 
\\ 
L1024.2   & 3.47 &  10793 &  0.0073  &   0.209   &  23457 & 0.0899\\ 
L1024.4   & 2.63 &  13225 &  0.0266  &   0.316   &  15216 &  0.0755\\ 
L1024.8   & 2.00 &  6718 &  0.0316  &   0.436    &  5538 & 0.0661\\ 
\\
SDSS.2   & 0.209  & 2198 & 0.0275 &     &  - &  \\
SDSS.4   & 1.38  & 1820 & 0.0529 &     &  - &  \\
SDSS.8   & 1.82  &  834 & 0.0640 &     &  - &  \\
\label{Tab1}                         
\end{tabular} 
\tablefoot{
The columns in the Table are as follows:\\ 
\noindent column 1: sample name XXX.j, where XXX is for $\Lambda$CDM
model or SDSS; $j$ gives
the smoothing kernel radius $R_B =j$  in \Mpc;\\ 
\noindent column 2: percolating threshold density for clusters, $P_C$, 
in mean density units;\\ 
\noindent column 3: number of clusters, $N_C$;\\  
\noindent column 4: filling factor of the largest cluster, $F_C$, at the
percolating threshold density;\\
\noindent column 5: percolating threshold density for voids, $P_V$, in
mean density units;\\ 
\noindent column 6: number of voids, $N_V$;\\
\noindent column 7: filling factor of the largest void, $F_V$, 
at this threshold density.
\\
Numbers of clusters and voids for the model L100 are found using
limiting system volumes $N_{lim} =200$, for other DM models using
$N_{lim} =50$, and for SDSS samples using $N_{lim} =500$.
First data block  presents percolation parameters of the DM 
model L512 for epochs at redshifts $z=30,~10,~3,~1,~0$,  smoothing
scale 1~\Mpc, and limiting volume $N_{lim} =500$.
}
\end{table} 
}

Table~\ref{Tab1} and Fig.~\ref{fig:L512evol} shows how percolation
functions change during the evolution of the cosmic web, and obtain
the form at the present epoch.  At the early epoch $z=30$ there are no
voids at threshold densities $D_t \le 0.8$, and no clusters at
$D_t \ge 1.2$. As the evolution proceeds, the interval of threshold
densities, where clusters and voids exist, increases. At early epochs
percolation functions of clusters and voids are rather symmetrical in
logarithmic scale.  This symmetry is gradually lost during the
evolution.

One measure of the connectivity of the web is the percolation
threshold density, which is found for both clusters and voids, $P_C$
and $P_V$.  Table~\ref{Tab1} and Fig.~\ref{fig:L512evol} show the
change of the percolation threshold density of clusters and voids with
the evolution epoch, $z$.  At the early epoch $z=30$ the distribution
of densities is almost Gaussian and symmetric around the mean density,
$D_t =1.0$.  Percolation functions at early epochs are very close to
respective functions for purely random samples \citep{Einasto:1986a}.
Thus the change of percolation threshold density with epoch describes
the growth of departures from the initial Gaussian density field to
its present non-Gaussian form.

The spread of densities around the mean density $D =1.0$ is at early
epochs proportional to the amplitude of density perturbations.  At the
recombination epoch, $z \approx 1000$, the amplitude of density
perturbations is of the same order, $\delta = D -1.0 \approx 10^{-3}$.  The
departure of percolation density threshold for clusters from the mean
density value, $D=1.0$, is of the same order,
$P_C \approx 1+ 10^{-3}$.  Similarly, the departure of percolation
density threshold for voids from unity is, $P_V \approx 1 - 10^{-3}$.
Both for clusters and voids the limiting values of percolation density
thresholds of random Gaussian samples are very close to 1.0.  We
accept as the percolation density threshold for random samples the
geometric mean of percolation threshold density for $z=30$ and
$z\approx 1000$: $P_{C0} =1.06 \pm 0.03$ for clusters, and
$P_{V0} = 0.94 \pm 0.03$ for voids, or approximately
$P_0 =1.00 \pm 0.05$ for both.  The departure of percolating threshold
densities from these values can be used as a measure of the departure
of the density field from a Gaussian one.

\begin{figure*}[ht]
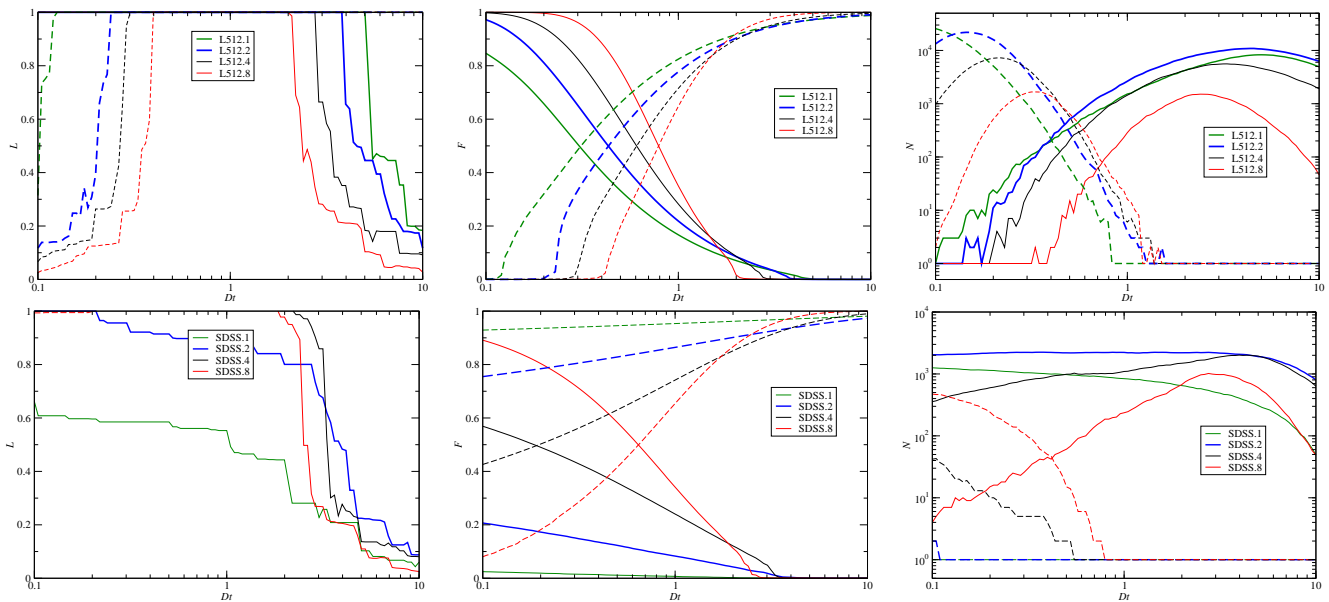
 
\centering 
\hspace{2mm}  
\resizebox{0.30\textwidth}{!}{\includegraphics*{L512_Lmax-D0_4.eps}}
\hspace{2mm}  
\resizebox{0.30\textwidth}{!}{\includegraphics*{L512_Vmax-D0_4.eps}}
\hspace{2mm}  
\resizebox{0.30\textwidth}{!}{\includegraphics*{L512_N-D0_4.eps}}\\
\hspace{2mm}  
\resizebox{0.30\textwidth}{!}{\includegraphics*{dr8_Lmax-D0_3.eps}}
\hspace{2mm} 
\resizebox{0.30\textwidth}{!}{\includegraphics*{dr8_Vmax-D0_3.eps}}
\hspace{2mm} 
\resizebox{0.30\textwidth}{!}{\includegraphics*{dr8_N-D0_3.eps}}
 \hspace{2mm}
\\ 
\caption{Percolation functions of model and observational samples:
  {\em Left:} lengths of largest clusters and voids,
  $\mathcal{L}(D_t) = L_{\mathrm{max}}/L_0$; {\em Center:} filling
  factors of largest clusters and voids,
  $\mathcal{F}(D_t) = V_{\mathrm{max}}/V_0$; {\em Right:} numbers of
  clusters and voids, $\mathcal{N}(D_t)$, as functions of the
  threshold density, $D_t$.  Lengths and filling factors are expressed
  in units of the total lengths and volumes of samples.  Top panels
  are for DM model L512, bottom panels are for SDSS samples.
  Percolation functions are found using smoothing kernels of radii 1,
  2, 4, and 8~\Mpc.  Solid lines show data on clusters, dashed lines
  on voids. Indices show the smoothing kernel length in \Mpc. Number
  functions $\mathcal{N}(D_t)$ of DM models in this Figure correspond
  to small system excursion limit $N_{\mathrm{lim}}=50$ cells. }
\label{fig:L512-dr8} 
\end{figure*}

\subsection{Percolation functions of DM  clusters and voids}

Consider first percolation functions of DM model clusters, i.e.  high-density
regions above the threshold density $D_t$ in our simulations, plotted
on Fig.~\ref{fig:L512-dr8}.  At very high threshold density there
exists only a few high-density regions --- peaks of ordinary clusters
of galaxies.  These peaks are isolated from each other, they cover a
small filling factor in space.  When we lower the threshold density,
the number of clusters increases, as well as the filling factor of the
largest cluster.  At certain threshold density, $D_t \approx 5$
(depending on the smoothing scale and the size of the computational
box), the number of clusters reaches a maximum.  At this threshold
density  large clusters still covers a low filling factor, and have 
 lengths, less than the size of the sample.  Most large clusters have
the form of conventional superclusters, consisting of high-density
knots, joined by filaments to a single system.  

When the threshold density still decreases, the length of the largest
cluster increases very rapidly, supercluster-like systems merge, and
at threshold $D_t = P_C$ the largest cluster reaches opposite
sidewalls of the model.  Since our models are periodic boxes, this
means that the cluster is actually infinite in length.  The
percolation threshold depends on the box size and on the smoothing
length. We shall investigate this dependence in more detail below.
Data for just percolating clusters and voids are given in
Table~\ref{Tab1}: percolating threshold densities $P_C$ and $P_V$,
numbers of clusters and voids, $N_C$ and $N_V$, and filling factors at
percolating threshold densities, $F_C$ and $F_V$.

Now consider percolation functions of voids.  At very small threshold
density, $D_t \ll 0.1$, there are no voids at all. Voids appear at a
certain threshold density level, depending on the smoothing kernel,
and their number rapidly increases with increasing $D_t$.  At very low
threshold density void sizes are small, they form isolated bubbles
inside the large over-density cluster, and the filling factor of the
largest void is very small. Voids in the density field have at the
smallest threshold density the length of the largest void,
$\mathcal{L}_V(D_t)  \le 0.5$, depending on the
smoothing scale and the size of the model.  Void bubbles are separated
from each other by DM sheets.  Some sheets have tunnels which  allows to
form some larger connected voids.  With increasing threshold density
the role of tunnels rapidly increases, tunnels join neighbouring
voids.  At certain threshold density ($D_t \approx 0.2$ for small
smoothing lengths) the largest void is percolating, but still not
filling a large fraction of the volume.  When we use larger smoothing
lengths then expanded high-density regions block tunnels between voids
at small threshold density, and percolation occurs at higher $D_t$.

The number of voids has a maximum at threshold density
$D_t \approx 0.1$ for the density field of smoothing scale
1~\Mpc. With increasing smoothing scale the maximum shifts to larger
$D_t$.  For $D_t \ge 1$ there is only one large percolating void, its
filling factor increases with the increase of $D_t$.

\subsection{The influence of smoothing scale to describe the cosmic web}

{ According to the presently accepted $\Lambda$CDM paradigm, there
  exist density fluctuations of all scales. This leads to fractal
  nature of the distribution of dark matter and galaxies with a
  transition to homogeneity on large scales
  \citep{Mandelbrot:1982uq, Jones:1988pm, Einasto:1993af}.
  The hierarchical character of dark matter distribution is well seen
  in voids with sub-voids, sub-sub-voids
  etc. \citep{Aragon-Calvo:2010ve, Aragon-Calvo:2010wd,
    Aragon-Calvo:2013}.  The fractal structure has no preferred
  scales.  However, physical processes on different scales are
  different.  On small scales inside halos non-gravitational
  (hydrodynamical) processes are dominant. This is a topic of special
  studies, where special questions are asked, such as galaxy formation
  \citep{White:1978}, galaxy evolution \citep{Tinsley:1968}, the
  number of galactic satellites \citep{Klypin:1999aa}, and many more.
  On larger scales purely gravitational processes are dominant.  We can ask:
  At which scale the transition from non-gravitational to
  gravitational character of processes occurs?}

Within halos dark and baryonic matters are separated.  Luminous matter
forms visible populations of main galaxies and satellite galaxies.  A
fraction of baryonic matter within halos is in the form of diffuse hot
coronas of main galaxies.  Using catalogs of luminous galaxies and
applying appropriate smoothing it is possible to restore approximately
the distribution of baryonic matter for comparison with smooth
distribution of dark matter.

Radii of DM halos can be estimated using visible objects, such as
satellites around giant galaxies, and other members of
clusters/groups.  Already early estimates have shown that radii of DM
halos around galaxies are of the order of 1~\Mpc\
\citep{Einasto:1974fv}, confirmed by recent observations of velocities
of galaxies in the nearby volume of space: satellites of giant
galaxies have orbital velocities up to a distance $\approx 1$~\Mpc\
from the central galaxy.  At larger distance the smooth Hubble flow
dominates \citep{Karachentsev:2002b}, showing the transition from DM
dominated halos to filaments.  This limit corresponds to a smoothing
kernel radius $\approx 0.6$~\Mpc.

\citet{Tempel:2014vn} searched for filaments in SDSS main galaxy
survey, and found that characteristic radii of galaxy filaments are
0.5~\Mpc. Authors showed that filaments of such radius have the
strongest impact on galaxy evolution parameters. Actually the radius
of filaments can be a bit larger, of the same order as the size of DM
halos of bright galaxies.

{ We chose 1~\Mpc\ as the scale of transition from dominantly
  non-gravitational to gravitational character of processes.  We
  consider the structure on smaller scales as the topic of galactic halos, the
  structure on larger scales the topic of the cosmic web, and
  smoothing with kernel length $R_B=1$~\Mpc\ as representing the true
  density field of gravitating matter of the web.  We use smoothing
  with larger kernels for methodical purposes to understand properties
  of the web on various scales, as done among others by
  \citet{Aragon-Calvo:2007, Cautun:2013aa, Cautun:2014qy}.
  Percolation functions, calculated for density fields with smoothing
  scales 1 and 2~\Mpc\ are plotted in Fig.~\ref{fig:L512-dr8} by bold
  lines; the scale 2~\Mpc\ is available in all our DM models.  $B_3$
  kernels $R_B=1,~2$~\Mpc\ correspond to Gaussian kernels
  $R_G = 0.6, ~1.2$~\Mpc.  }

The contrast in the behaviour of clusters and voids is the largest
when we use original density fields of models.  
Smoothing shifts part of DM from high-density regions to their
surrounding,   increases filling factors of clusters, and
decreases filling factors of voids, especially on small and medium
threshold densities.  In this way smoothing decreases density
contrast, which leads to the decrease of percolation thresholds of
clusters, and to the increase of percolation thresholds of voids.

\subsection{Comparison of models of different size}

Table~\ref{Tab1} shows clearly that percolation functions of DM
models of different size are very similar to each other for identical
smoothing lengths.  We use this similarity to define percolation
parameters of samples.  One of the principal geometrical properties of
the cosmic web is the connectivity of over- and under-density regions,
or clusters and voids in our terminology.  The connectivity can be
measured by the percolation threshold density of clusters and voids.

\begin{figure}[ht] 
\centering 
\hspace{2mm}  
 \resizebox{0.45\textwidth}{!}{\includegraphics*{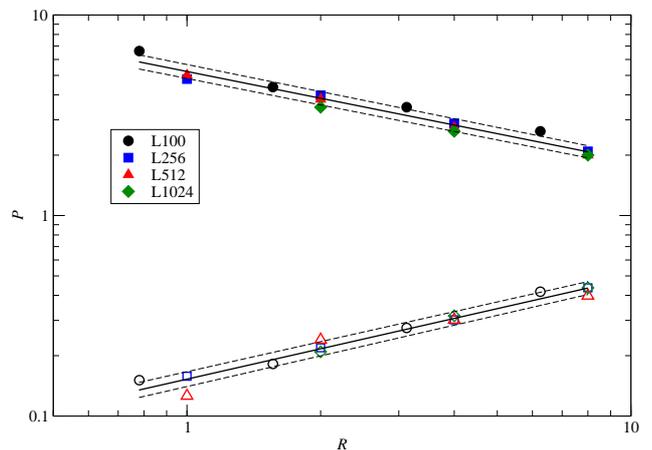}}
\\ 
\caption{Percolation threshold of clusters and voids, $P_C$ and $P_V$,
  of models L100, L256, L512, L1024 for various smoothing lengths
  $R_B$.  { Filled symbols are for clusters, open symbols for
    voids. Linear fits according to eqn.~(\ref{pp}) are shown by solid
    lines, 95~\% confidence limits by dashed lines.}}
\label{fig:perc} 
\end{figure}

In Fig.~\ref{fig:perc} we show percolation threshold densities as
functions of the smoothing kernel length,  $R_B$.  Data are given for
all our models and smoothing lengths.  The Figure shows that there
exists a very close relationship of percolation threshold densities
and the smoothing kernel length, both for clusters and voids.  The
relationship between $P_C$ and  $R_B$ is very close, and almost linear
in log-log representation:
\begin{equation}
\log P_C =  a_C +  b_C \times \log R_B,
\label{pp}
\end{equation}
and a similar equation for void percolation threshold densities, $P_V$
and $R_B$.  Constants of equations have values
$a_C = 0.718 \pm 0.014$, $b_C = -0.444 \pm 0.025$ for clusters, and
$a_V = -0.816 \pm 0.015$, $b_V = 0.503 \pm 0.027$ for voids.  The
scatter of individual values from the mean relationship is rather
small.  At $R_B=1$ the percolation threshold of clusters is
$\log P_C(1) = 0.718 \pm 0.014$, or $P_C(1)= 5.23 \pm 0.31$, and of
voids is $\log P_V(1) = -0.816 \pm 0.015$, or
$P_V(1)=0.152 \pm 0.012$.  The deviation from the percolation
threshold density from a random distribution with
$\log P_0 = 0.00 \pm 0.05$ (or more exactly $P_{C0}=1.06 \pm 0.03$ and
$P_{V0}=0.94 \pm 0.03$) is very clear, and exceeds the mean scatter of
values around the mean relationship by a large margin.

The relationship between $P_C$ and $R_B$ (and between $P_V$ and $R_B$)
is the same for all models, in spite of large differences of sizes of
models from $L=100$ to $L=1024$~\Mpc.  Actually the relationship
(\ref{pp}) is valid ever in a broader scale. \citet{Einasto:1986a}
found $\log P_C = 0.55$ and $\log P_V = -0.70$, for a $\Lambda$CDM
model of box length 40~\Mpc, using smoothing scale
$\approx 1.25$~\Mpc, not far from our present results.  This means
that the percolation threshold density is a well-defined and stable
characteristics of DM model samples.

\begin{figure}[ht] 
\centering 
\hspace{2mm}  
 \resizebox{0.45\textwidth}{!}{\includegraphics*{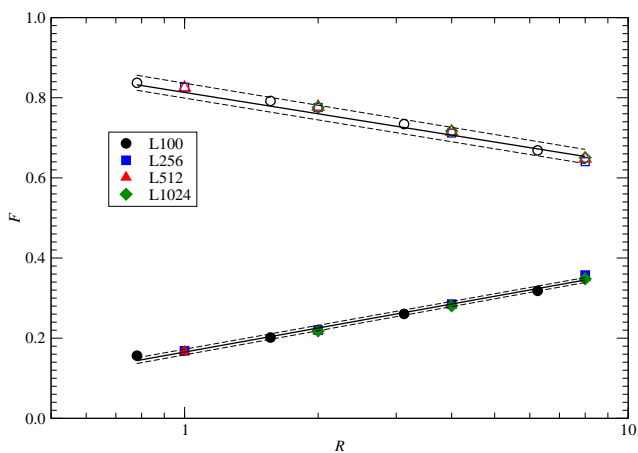}}
\\ 
\caption{Filling factors of largest clusters and voids at mean
  threshold density, $\mathcal{F}_C(1)$ and $\mathcal{F}_V(1)$, of
  models L100, L256, L512, L1024 for various smoothing lengths
  $R_B$. { Filled symbols are for clusters, open symbols for
    voids. Linear fits according to eqn.~(\ref{ff2}) are shown by solid
  lines, 95~\% confidence limits by dashed lines.}}
\label{fig:ff} 
\end{figure}

Filling factors of largest clusters and voids at mean threshold
density, $D_t=1.0$, depend on the smoothing kernel size, as seen in
Fig.~\ref{fig:ff}.  We can express this relationship as follows:
\begin{equation}
\mathcal{F}_C(R_B) =  f_C +  g_C \times \log R_B,
\label{ff2}
\end{equation}
and a similar equation for voids.  We get for constants of the
equation values: $f_C = 0.1652 \pm 0.0028$, $g_C = 0.1992 \pm 0.0051$;
and for voids $f_V = 0.8286 \pm 0.0028$, $g_V = -0.1962 \pm 0.0050$.
From these constants we get
$\mathcal{F}_C(1) + \mathcal{F}_V(1) = 0.9938$, the summed filling
factors of largest clusters and voids at smoothing length $R_B=1.0$.  It
is a bit less than unity, since small isolated clusters/voids except
the largest ones have very small volumes.  We see that filling factors
of largest clusters and voids at mean threshold densities of all our
models fit the same relationship very accurately.

If the relationship (\ref{ff2}) is valid for larger smoothing kernels, then at
$R_B \approx 30$~\Mpc\ filling factors of clusters and voids at mean
threshold densities become equal, and at still larger kernels cluster
filling factors exceed filling factors of voids.  The kernel
$R_B \approx 30$~\Mpc\  corresponds to Gaussian kernel $R_G \approx
18$~\Mpc.  

\subsection{Total filling factors}

\begin{figure}[ht]
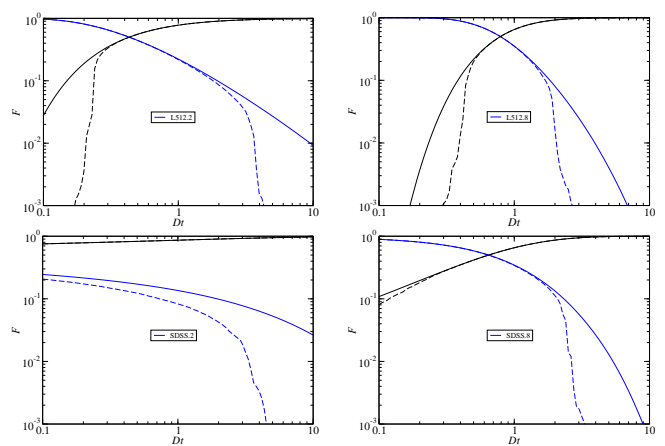
 
\centering 
\hspace{2mm}  
\resizebox{0.22\textwidth}{!}{\includegraphics*{L512.2_Ftot-Fmax_D0_B.eps}}
\hspace{2mm}  
\resizebox{0.22\textwidth}{!}{\includegraphics*{L512.8_Ftot-Fmax_D0_B.eps}}\\
\hspace{2mm}  
\resizebox{0.22\textwidth}{!}{\includegraphics*{SDSS.2_Ftot-Fmax_D0_B.eps}}
\hspace{2mm}  
\resizebox{0.22\textwidth}{!}{\includegraphics*{SDSS.8_Ftot-Fmax_D0_B.eps}}\\
\caption{Filling factor functions of the model L512:
  total filling factors of clusters and voids,
  $\mathcal{F}_{\mathrm{tot}}$, and filling factors of maximal
  clusters and voids, $\mathcal{F}_{\mathrm{max}}$, as functions of
  threshold density, $D_t$, plotted in logarithmic scales. Solid lines
  are for total filling factors of high- and low-density regions,
  dashed lines for maximal clusters and voids; blue colour marks
  clusters and  black colour voids. Functions are calculated using
  smoothing kernels 2 and 8~\Mpc\ (left and right panels).  Bottom
  row: the same functions for observed SDSS samples.  }
\label{fig:F_Fmax_D0} 
\end{figure}

So far we have used only filling factors of largest clusters and voids
to characterise properties of the web. Now we discuss also the
relationship between total filling factors, and filling factors of
largest systems.  We show in Fig.~\ref{fig:F_Fmax_D0} total filling
factors of high- and low-density regions,
$\mathcal{F}_{\mathrm{tot}}$, and filling factors for maximal clusters
and voids, $\mathcal{F}_{\mathrm{max}}$, as functions of threshold
density, $D_t$.  { Upper panel shows filling factor functions for
  the L512 model; functions for other models are very similar. Lower
  panel shows similar functions for SDSS samples.  Left and right
  panels show filling factor functions applying smoothing kernel radii
  $R_B =2,~8$~\Mpc, respectively.  }

Filling factor functions for
models of different box sizes but identical smoothing kernels are
rather similar, both for clusters and for voids.  As expected, the
size of the smoothing kernel has a large effect to filling factor
functions.  At threshold density $D_t = 10$ models of all sizes and
smoothing kernel $R_B =2$~\Mpc\ have total filling factor of clusters
$\mathcal{F}_{\mathrm{tot}} \approx 10^{-2}$.  At threshold level
$D_t=0.1$ and the same smoothing kernel the total filling factors of
voids are $\mathcal{F}_{\mathrm{tot}} \approx 0.03$, also for models
of all sizes.  Larger smoothing with kernel $R_B =8$~\Mpc\ decreases
the threshold density of clusters at the same filling factor to
$D_t \approx 5$, and increases the threshold density of voids to
$D_t \approx 0.3$.

At small threshold density, $D_t \le 1$, DM samples have
$\mathcal{F}_{\mathrm{max}} \approx \mathcal{F}_{\mathrm{tot}}$, i.e. the
largest cluster fills the whole over-density volume.  In other words,
the summed volume of all clusters except the largest one is small or
zero.  When we compare these functions at higher threshold densities,
we see that the filling factor of the largest cluster decreases
rapidly with increasing $D_t$, and total filling factors of smaller
clusters increases, i.e. curves for $\mathcal{F}_{\mathrm{tot}}$ and
$\mathcal{F}_{\mathrm{max}}$ diverge.

The behaviour of filling factor functions for voids is opposite.  For
large threshold densities the filling factors of the largest void are
equal to the total filling factors --- there exists only one large
void.  At lower threshold densities the fraction of small voids
increases, and functions for $\mathcal{F}_{\mathrm{tot}}$ and
$\mathcal{F}_{\mathrm{max}}$ diverge.

The Table~\ref{Tab1} shows that filling factors of largest
clusters/voids at percolation levels, $\mathcal{F}_C(D_t=P)$, as
functions of the smoothing scales of models, $R_B$, have a larger
scatter, than the relationship (\ref{ff2}).   This means that there exists
small differences between models of different size.  For small
smoothing lengths ($R_B=1,~2$) filling factors of largest DM clusters
at percolating threshold is $\mathcal{F}_C(P) \approx 10^{-2}$, and of
largest DM voids $\mathcal{F}_V(P) \approx 5\times 10^{-2}$, for
models of all sizes.

\citet{Sahni:1997ai} calculated percolation functions in the form
$v_{\mathrm{max}} = F_{\mathrm{max}}/F_{\mathrm{tot}}$, and plotted
them as functions of $F_{\mathrm{tot}}$, and of density contrast,
$\delta_t = 1 -D_t$, both for clusters and for voids. These
representations are complementary to representations, presented in
Figs.~\ref{fig:L512-dr8} and \ref{fig:F_Fmax_D0}.  We prefer to have
as argument in percolation functions the threshold density level.

\subsection{Topological properties of the cosmic web}

Percolation functions are not meant to describe topological properties
of the cosmic web in such way as genus and Minkowski functional
approaches allow. However, our data are sufficient to make distinction
between main types of topology: cellular or Swiss-cheese type,
sponge-like, and meatball-like.  The distinction between these
topologies is given by the percolating threshold densities of clusters
and voids. Cellular topology corresponds to the case when clusters are
percolating, but voids not.  If both clusters and voids are
percolating then we have the sponge topology.  When voids are
percolating, but clusters not, we have the meatball-like topology.

Table~\ref{Tab1} shows percolating threshold densities $P= D_t$ for 
our samples.  We see that   DM models have all three
types of topology: cellular at small $P$, sponge-type at medium
$P$, and meatball-type at large $P$.  Limits for the sponge
topology are the broadest for the smallest smoothing length, 1~\Mpc. 
Voids of SDSS samples are always percolating, thus at small and medium
threshold density until the percolation of clusters, $P \approx 2$,
the topology is of sponge type, and at larger threshold density of
meatball-type.

\section{Percolation properties of model and SDSS
  samples}  

The focus of the discussion in this Section is the comparison of
properties of the real luminosity density field and the simulated dark
matter density field.  Also we study the influence of observational
selection effects to percolation properties. 

\subsection{Percolation properties of SDSS clusters and voids}

Lower panels of Fig.~\ref{fig:L512-dr8} show percolation functions of
SDSS clusters and voids, lower panels of Fig.~\ref{fig:F_Fmax_D0}
present SDSS filling factor functions.  The comparison of percolation
functions of DM models and SDSS samples shows the presence of
important differences.

The major difference between models and observations is the absence in
SDSS samples of fine structure of voids. At all smoothing lengths SDSS
voids are percolating, and the percolation threshold density is not
defined.  For small smoothing lengths the percolating SDSS void is the
only void.  As the smoothing scale increases there appear additional
small SDSS voids at low threshold densities.  The total number of voids,
$\mathcal{N}_V$, increases with increasing smoothing length.  These
isolated small voids are artificial, and are created by blocking
tunnels between sub-voids with increasing sizes of clusters by
smoothing.  At low $D_t$ the largest SDSS void forms a filling factor
$\mathcal{F}_V  \approx 0.1$  for small
smoothing kernel.  With increasing $D_t$ the filling factor of voids
rapidly increases with $D_t$, and reaches a value
$\mathcal{F}_V \approx 0.98$ at the highest $D_t$.

\begin{figure*}[ht] 
\centering 
\resizebox{0.98\textwidth}{!}{\includegraphics*{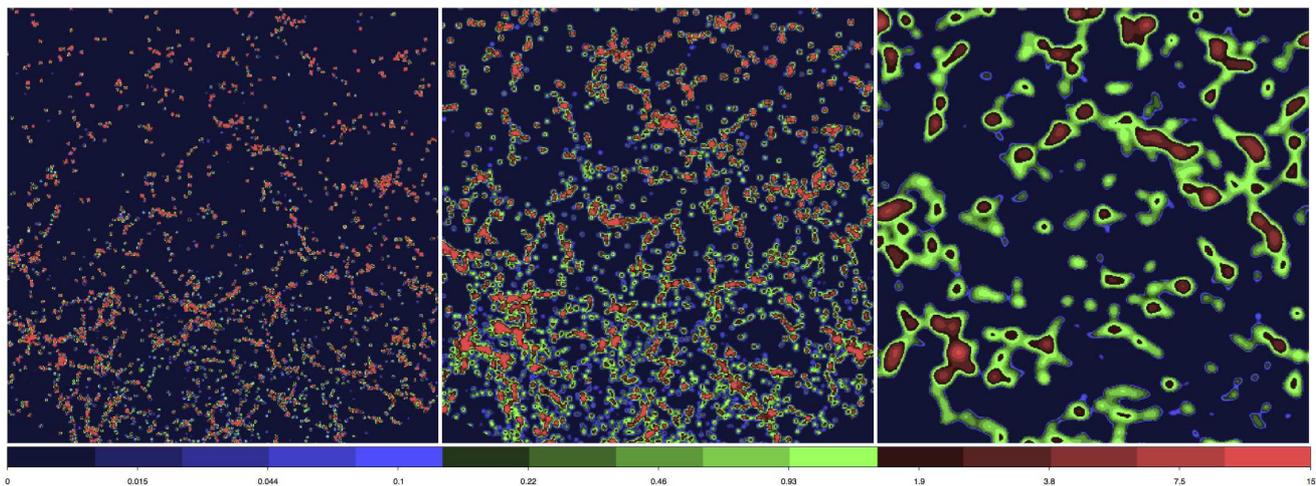}}
\hspace{2mm}  
\caption{Luminosity density fields { in
    $400\times 400\times 1$~\Mpc\ slices at identical $z-$coordinates
    of the central region of the SDSS sample, smoothed with kernels of
    radii 1, 2 and 8~\Mpc\ (left, middle and right panels). This
    Figure illustrates the effect of the smoothing length to
    geometrical properties of clusters and voids. Densities are
    expressed in logarithmic scale in interval 0.005 to 15 in mean
    density units}.  The colour code is identical in all panels.  }
\label{fig:sdssdens} 
\end{figure*}

\begin{figure*}[ht] 
\centering 
\resizebox{0.98\textwidth}{!}{\includegraphics*{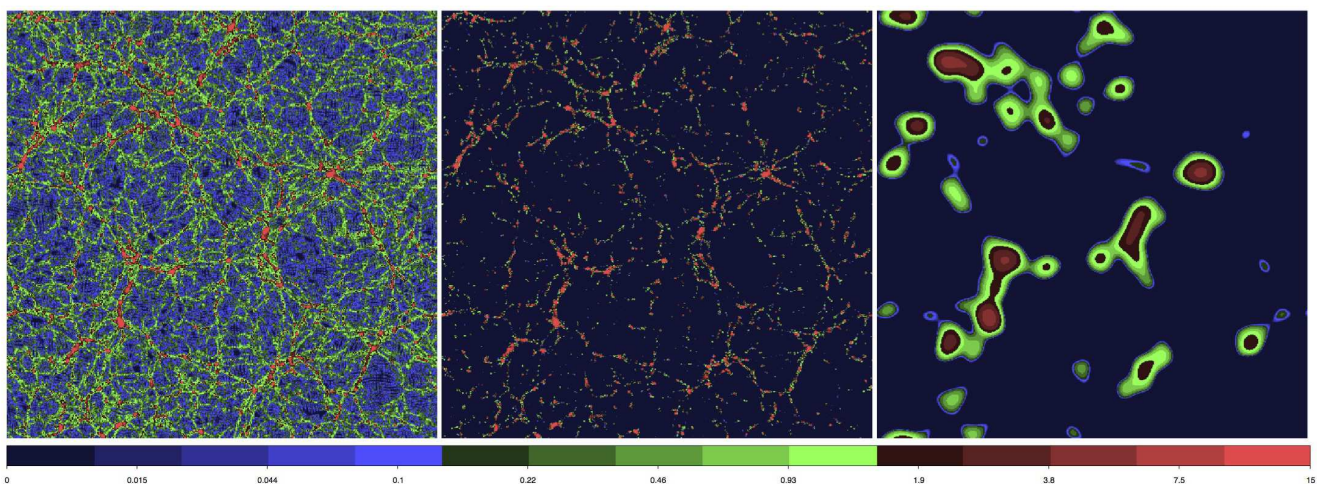}}
\caption{Density fields { in $256\times 256\times 0.5$ \Mpc\ slices
    of the L256 model at identical $z-$coordinates.  The left panel
    shows the density field smoothed with the kernel of radius 1~\Mpc\
    in density interval 0.005 to 15 in mean density units; the colour
    code below corresponds to this field}.  The middle panel shows the
  same density field in density interval 1.5 to 15 in mean density
  units. Here faint filaments between high-density knots are
  invisible. The right panel shows the field smoothed with kernel
  8~\Mpc\ in interval 1.5 to 15 in mean density units. Filaments are
  thicker and percolate easier. { In all panels densities are
    expressed in logarithmic scale.}  }
\label{fig:L256dens} 
\end{figure*}

There are also differences between properties of clusters of models
and observations.  For smoothing length $R_B=1$~\Mpc\ SDSS cluster
samples do not percolate at all, and the length of the largest cluster
is smaller than the sample size, $\mathcal{L}_C(0.1) = 0.6$.  At this
smoothing length clusters form isolated systems due to the absence or
weakness of galaxy filaments between high-density knots.  
{ The difficulty to trace observationally  thin filaments 
 and  sheets has been pointed out also by \citet{Cautun:2014qy}. } To
illustrate this phenomenon we show in Fig.~\ref{fig:sdssdens} the
luminosity density fields of the SDSS sample, smoothed with various
smoothing kernels. In the left panel of Fig.~\ref{fig:sdssdens} the
SDSS density field is smoothed with kernel 1~\Mpc, in the middle panel
with kernel 2~\Mpc. We see that larger smoothing increases the volume
of faint knots, located between high-density knots, and helps clusters
to percolate. This effect is also well seen in DM model sample in
Fig.~\ref{fig:L256dens}, where we show density fields of the model
L256 for smoothing lengths 1 and 8~\Mpc, and in different density
intervals.

The percolation of SDSS samples occurs at lower threshold density than
expected from the comparison with DM model samples.  In comparison to
DM clusters, filling factor functions of SDSS clusters are shifted ---
cluster filling factors are lower, and void filling factors higher.
At lowest threshold densities filling factors of DM clusters are close
to 1 for all smoothing lengths. In contrast, filling factors of SDSS
clusters are much lower, only $\mathcal{F}_C(0.1)=0.025$ for the
smoothing kernel $R_B=1$.

It is remarkable that the number of L512 and SDSS clusters at
smoothing length $R_B = 8$~\Mpc\ reaches maximal values at
$D_t \approx 5$, and that the number of clusters is also approximately
the same, for identical small system elimination threshold
$N_{lim} = 500$.  The smoothing length $R_B = 8$~\Mpc, and the
threshold density $D_t \approx 5$ are often used to find superclusters
of galaxies \citep{Liivamagi:2012aa}. { Samples L512 and SDSS have
  approximately the same total volumes, thus the close number of
  superclusters in both samples shows that the L512 model represents
  the real cosmic web on supercluster level very well.}

\subsection{Isolated  clusters and the separation of
  intrinsic and selection effects}

A further difference of DM and SDSS samples lies in number functions
of clusters for various smoothing lengths.  Fig.~\ref{fig:L512-dr8}
shows that, for smoothing lengths 1, ~2,~4~\Mpc\ and $D_t \le 4$, the
number of clusters of SDSS samples is almost independent on the
threshold density.  Only the sample SDSS.8 has a cluster number
function, similar to the number functions of DM samples:  small
number of clusters at low threshold density, and increasing number
with increasing $D_t$.

At very low threshold density, $D_t =0.1$, the DM samples have one
large cluster ($\mathcal{N}_C=1$), filling almost the whole space,
$\mathcal{F}_C(0.1) \approx 1$, see Fig.~\ref{fig:L512-dr8}.  In
Fig.~\ref{fig:L256dens} we show the high-resolution density field of
the L256.1 sample.  Left panel shows the density field, smoothed with
1~\Mpc\ kernel, and plotted { in logarithmic scale in density
  interval 0.005 to 15 in mean density units.}  The large DM cluster
contains numerous small and medium-sized knots.  Knots of the DM
density field are joined to a single high-density region (cluster) by
low-density DM filaments and sheets.  These faint DM filaments and
sheets isolate numerous small voids --- bubbles in the high-density
matter.  For this reason the number of voids in DM samples at low
threshold densities is high.

When we increase the threshold density to $D_t = 1.5$, shown in the
middle panel of Fig.~\ref{fig:L256dens}, then faintest DM filaments
between high-density knots became invisible.  At this density
threshold most small knots in the DM density field become isolated
clusters, and voids merge.  For this reason the number of high-density
systems (DM clusters) increases with the increase of the threshold
density, the number of voids in DM samples decreases, and lengths of
largest voids increase. { Notice that the general character of the
  high-resolution DM density field in the middle panel of
  Fig.~\ref{fig:L256dens} is very similar to the SDSS density field,
  shown in the left panel of Fig.~\ref{fig:sdssdens}.}

If we use the largest smoothing kernel, $R_B= 8$~\Mpc, shown in right
panels of Figs.~\ref{fig:sdssdens} and \ref{fig:L256dens}, then many
previously small clusters join to larger clusters, and the number of
clusters decreases.  This effect is observed in DM models of all
sizes, and in the observational SDSS sample.  { The general
  character of DM and SDSS density fields at this smoothing kernel is
  also very similar.}

\begin{figure*}[ht]
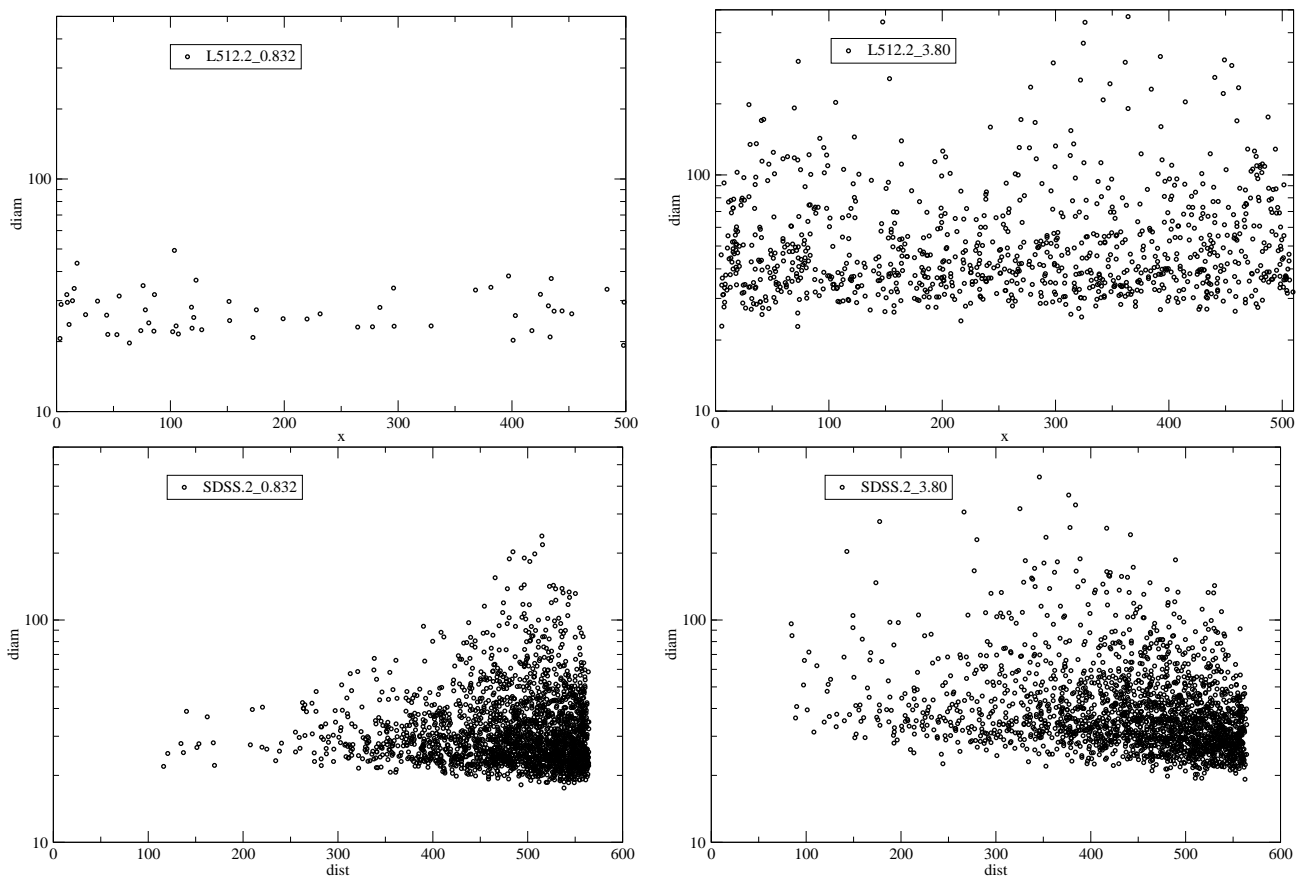
 
\centering 
\hspace{2mm}  
\resizebox{0.45\textwidth}{!}{\includegraphics*{L512.2_diam-x_0.8318.eps}}
\hspace{2mm}  
\resizebox{0.45\textwidth}{!}{\includegraphics*{L512.2_diam-x_3.8.eps}}\\
\hspace{2mm}  
\resizebox{0.45\textwidth}{!}{\includegraphics*{L_distr_SDSS.2_0.8318.eps}}
\hspace{2mm}  
\resizebox{0.45\textwidth}{!}{\includegraphics*{L_distr_SDSS.2_3.80.eps}}
\hspace{2mm}  
\\
\caption{{\em Top:} diameters of clusters vs. $x$-coordinate of the
  sample L512.2, at threshold densities $D_t=~0.832,~3.80$, {  left
  and right panels, respectively}.  {\em
    Bottom:} diameters of clusters vs. distance from the observer for
  the SDSS.2 sample at the same threshold densities.  { Diameters
    and distances ($x$-coordinates) are given in \Mpc, diameters are
  plotted in a logarithmic scale. } Largest percolating clusters are
  not shown. In all samples a small cluster exclusion limit,
  $N_{lim}=500$, is applied.}
\label{fig:diam} 
\end{figure*}

To understand better the nature is this effects let us compare numbers
and diameters of clusters of the DM model sample L512.2 with numbers
and diameters of clusters of the observed sample SDSS.2.  Diameters of
individual clusters as functions of the distance from the observer are
shown in Fig.~\ref{fig:diam}, for the threshold density $D_t=0.832$,
and the percolating threshold density of the L512.2 sample,
$D_t=3.80$. Upper panels are for the model sample L512.2, lower panels
for the observed sample SDSS.2.  

At very low threshold densities the model sample L512.2 has no
isolated clusters: the whole over-density region contains one
percolating cluster. At threshold density $D_t=0.832$ the sample
L512.2 has one large percolating cluster, and about 60 small isolated
clusters --- peaks of the DM density field in low-density regions.
These clusters are distributed evenly with distance, and have
diameters $\sim 25$~\Mpc.

The observed sample SDSS.2 has at low and medium threshold densities
an approximately constant cluster number function,
$\mathcal{N}_C(D_t)$, as seen in Fig.~\ref{fig:L512-dr8}.  The
distribution of cluster diameters with distance, shown in
Fig.~\ref{fig:diam} for $D_t=0.832$, is very similar at small and
medium threshold densities, $D_t \le 1$. At these threshold densities
SDSS clusters are isolated, and almost identical at different
threshold density levels, and for smoothing kernels $R_B =1,~2$~\Mpc;
compare left and middle panels of Fig.~\ref{fig:sdssdens}.  Cluster
positions and diameters fluctuate slightly due to inclusion of fainter
cluster envelopes by decreasing threshold density.  {
  Fig.~\ref{fig:diam} shows that the number of SDSS.2 clusters
  increases rapidly with the distance from the observer; maximal
  diameters also increase with distance.}  Partly this effect is due
to the conical geometry of the SDSS sample: the volume of the sample
increases with distance.  But the large number of clusters in SDSS
samples is mainly caused by the absence of faint galaxy filaments
joining high-density knots at low threshold density to a connected
system, see Fig.~\ref{fig:sdssdens}.  Moreover, the SDSS sample is
flux-limited, and at large distance fainter galaxies are not visible,
which causes a further increase of the number of large clusters at
greater distance from the observer.

Fig.~\ref{fig:diam} shows that at the percolating threshold density of
the L512.2 sample, $D_t=3.80$, the general trend of diameter
distributions of L512.2 and SDSS.2 samples is fairly similar, if we
ignore differences due to the conical shape of the SDSS.2 sample.  At
this threshold density almost all high-density regions of both samples
are considered as isolated clusters, and have thus a similar
character.

When we use the smoothing length 8~\Mpc, the number density function
$\mathcal{N}_C$ of the SDSS sample has a shape, similar to the shape
for the L512 sample, as seen in Fig.~\ref{fig:L512-dr8}.  This means,
that additional smoothing restores bridges between high-density knots
of the SDSS sample, which leads to the loss of most isolated clusters
at small threshold densities, see also right panel of
Fig.~\ref{fig:sdssdens}.

Void distributions at low threshold densities are  very different
in model and real samples.  Model samples have at these thresholds and
at small smoothing lengths numerous small isolated void bubbles, and
one percolating void.  The SDSS sample has only one percolating void,
and no small voids at all.

But notice that the change of the smoothing scale affects cluster
lengths and numbers of clusters in a different way.  Smoothing with
4~\Mpc\ kernel restores the filamentary character of the SDSS density
field (see the next subsection), as characterised by the length of the
largest cluster, but not the elimination of small isolated clusters.
This means that at the smoothing scale 4~\Mpc, and the threshold
density $D_t=0.832$, there exists one large percolating cluster, but
numerous small isolated clusters remain.  Larger smoothing joins these
isolated clusters to the dominating one.

The elimination of small isolated clusters at low threshold densities
with 8~\Mpc\ smoothing, but not with 4~\Mpc\ smoothing, suggests that
peaks of the SDSS density field are quasi-regularly located with
mutual distances of the same order, along the filamentary web.  If
peaks would be randomly located, then the decrease of the number of
isolated clusters would be more gradual.  An independent confirmation
of this result comes from the study by \citet{Joeveer:1978dz}, who
found that groups and clusters of galaxies of the Perseus-Pisces
supercluster form a long chain with mutual distance between
clusters/groups of $\sim 8$~\Mpc.  This result was confirmed by a
recent study of \citet{Tempel:2014fk}, who demonstrated that galaxy
filaments are as pearl necklaces where groups form density
enhancements at a mutual distance $\approx 7$~\Mpc\ from each other.

These differences in percolation functions and cluster diameter
distributions between model and observed samples are caused by two
separate effects.  The main effect is due to the absence of very faint
filamentary systems in the observed sample, present in DM models.  The
second effect is caused by selections: the SDSS sample is conical, and
at large distance from the observer faint galaxies are not included
into the sample, which makes faint galaxy filaments, connecting
high-density knots, invisible.

The analysis of the distribution of isolated clusters  shows
that percolation functions are sensitive not only to general
geometrical properties of the cosmic web, but also to the presence of
faint filaments between high-density knots, and to the regular
displacement of high-density knots of the web.  The analysis also
shows that it is relatively easy to separate intrinsic and selection
effects in percolation functions.

\subsection{Filamentary character of the cosmic web}

As seen in Fig.~\ref{fig:L512-dr8}, largest clusters of DM samples
have for threshold density $D_t \le P$ identical percolating lengths,
$\mathcal{L}_C(D_t) \equiv 1$, and a very {\em rapid} decrease of the length
with the increase of the threshold density at $D_t > P$.  This rapid
decrease of the length with increasing $D_t$ is characteristic in a
filamentary web.  Percolation threshold depends on the smoothing
length.  Smoothing decreases density contrast, thus for larger
smoothing lengths percolation occurs at lower threshold densities. In
all cases percolation occurs at thresholds, much larger than the
percolation threshold for a random sample, $P_{C0} = 1.06$.  Voids of
dark matter samples have similar behaviour.  At threshold densities
$D_t < P$ void lengths decrease rapidly with decreasing $D_t$.  DM
void percolation thresholds, $P_V$, are much lower than void
percolation thresholds for random samples, $P_{V0} = 0.94$.

\begin{figure*}
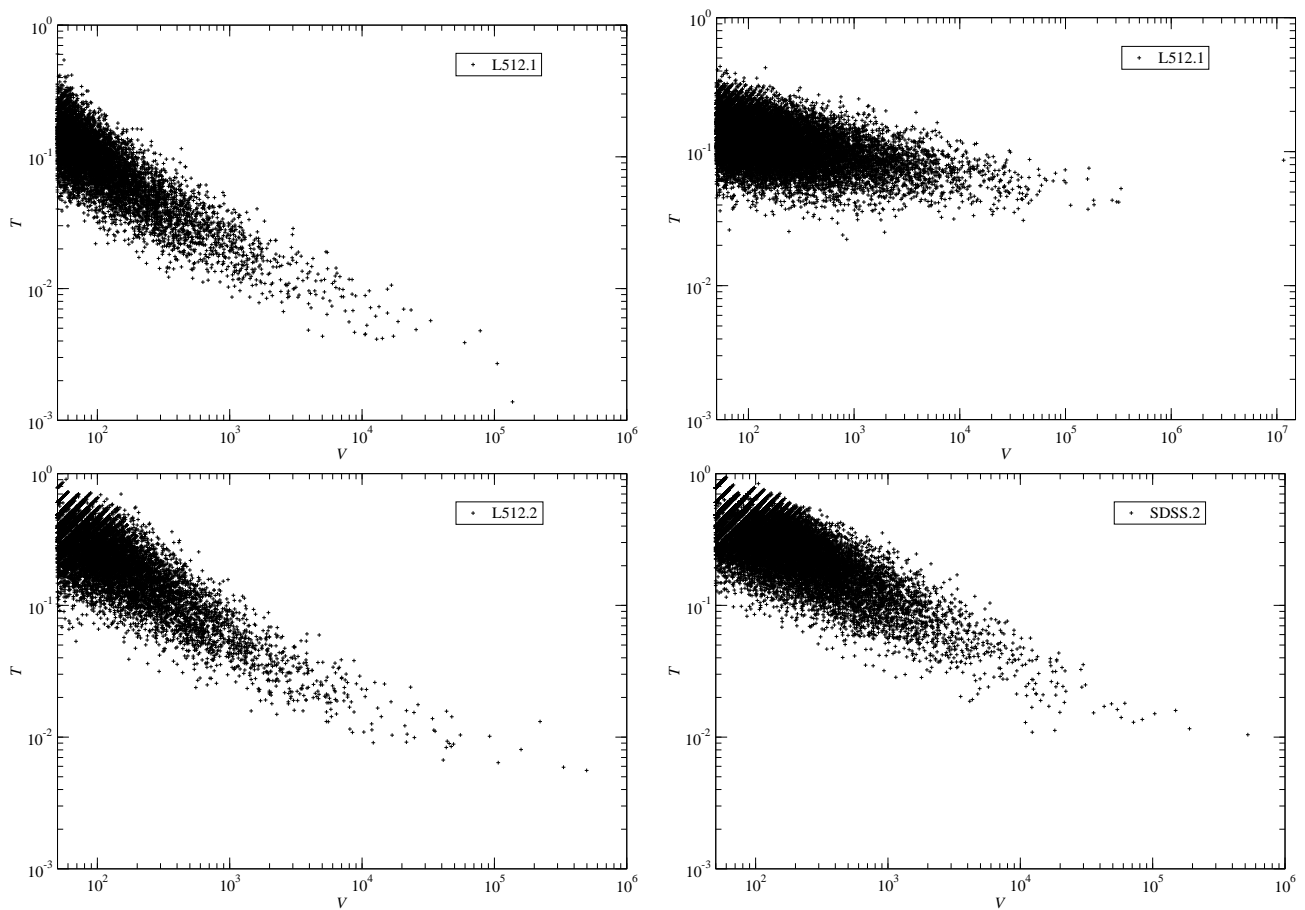

\centering 
\hspace{2mm}  
 \resizebox{0.45\textwidth}{!}{\includegraphics*{scl0_007_5.01_fat-vol.eps}}
\hspace{2mm}  
 \resizebox{0.45\textwidth}{!}{\includegraphics*{scl0_007_void_0.126_fat-vol.eps}}\\
\hspace{2mm}  
\resizebox{0.45\textwidth}{!}{\includegraphics*{scl1_008_3.8_50_fat-vol.eps}}
\hspace{2mm}  
\resizebox{0.45\textwidth}{!}{\includegraphics*{scl_dr8_a2_50_fat-vol.eps}}
\\
\caption{{ {\em Top} Fatness factors, $T = V/D_m^3$, of clusters
    (left panel) and voids (right panel) of the L512.1 sample at
    respective percolation thresholds, $D_t=5.01$ and $D_t=0.126$.
  {\em Bottom:} fatness factors of L512.2 clusters (left) and SDSS.2
  clusters (right) samples at the threshold density $D_t=3.80$.
  Fatness factors are given as functions of their volumes, $V$ (in
  cubic \Mpc), and were calculated using small system excursion limit
  $N_{\mathrm{lim}}=50$ cells.}}
\label{fig:fatness2} 
\end{figure*}

It should be noted, that the filamentary character is related to two
aspects of the distribution: a rapid change of the length of largest
clusters/voids with changing density threshold, {\em and} a deviation
of the respective threshold from the threshold for random samples.
The larger this deviation the more filamentary  the web is.  In a
random density field the length of largest clusters/voids also changes
rapidly with the change of the threshold density (see
Fig.~\ref{fig:L512evol}), but the threshold density difference
condition is not observed.

The behaviour of length functions of observed samples is different
from the behaviour of DM model samples.  To see differences between
model and observed samples, we compare in Fig.~\ref{fig:L512-dr8}
length functions of SDSS and L512 samples at various smoothing
lengths.  For smoothing kernel $R_B =1$~\Mpc\ the largest cluster of
the SDSS.1 sample at lowest threshold density $D_t = 0.1$ has a length
$\mathcal{L}_C \approx 0.6$, thus the SDSS.1 cluster sample does not
percolate at all, and the sample has a meat-ball type of the galaxy
distribution.  At larger threshold densities maximal length of
clusters decreases with increasing $D_t$ {\em slowly}.

For smoothing length $R_B =2$~\Mpc\ the largest cluster of the SDSS.2
sample percolates at the threshold density $P_C = 0.2$, much lower
than the percolation threshold density for random samples,
$P_{C0} =1.06$, and for the DM sample L512.2.  This means that at this
smoothing length the SDSS.2 sample is still mainly a sample of
isolated high-density regions, as seen in the middle panel of
Fig. ~\ref{fig:sdssdens}.  But the SDSS.2 sample has also a
differential luminosity selection effect. At smaller distance from the
observer fainter galaxies lie within the observational window of
apparent magnitudes, and a weak filamentary system of galaxies between
high-density knots is present, as seen from Fig.~\ref{fig:sdssdens}.
At larger distance from the observer this weak filamentary character
of the SDSS.2 sample breaks down.  This change of the character of the
web influences the length function.  The volume of the nearby region,
where percolation is easier, is much smaller than the volume of the
more distant region.  For this reason, in the sample SDSS.2 as a whole
a meat-ball type of the galaxy distribution dominates.

At smoothing lengths 4 and 8~\Mpc\ the behaviour of the SDSS cluster
length functions $\mathcal{L}_C(D_t)$ at large threshold density is
almost similar to the behaviour this function in the DM L512 sample
--- a rapid decrease of the cluster length with increasing threshold,
as seen in Fig.~\ref{fig:L512-dr8}.  However, near the percolation
threshold the change of the cluster length with changing $D_t$ is
slower than in a DM model of the same smoothing scale.  Thus the
percolation of SDSS samples occurs at lower threshold density than
expected from the DM sample with similar smoothing scales, L512.4 and
L512.8.  In model samples the rapid increase of the length of clusters
near the percolation threshold is fostered by the presence of
filaments near knots.  In the SDSS sample at large distance filaments
are weaker, thus a bit lower threshold density is needed to get the
maximal length of the cluster, equal to the characteristic length of
the sample, $L_0$.  If this deviation of the SDSS length function near
the percolation threshold is ignored, and the length functions of SDSS
samples are interpolated until the percolation threshold in a way,
similar to L512 samples, we get for the percolation threshold of
SDSS.4 and SDSS.8 values, very close to values for DM samples L512.4
and L512.8.  Fig.~\ref{fig:sdssdens} shows that at smoothing scale
8~\Mpc\ the filamentary character of the SDSS sample is practically
restored over the whole depth of the SDSS sample.

\begin{figure*}
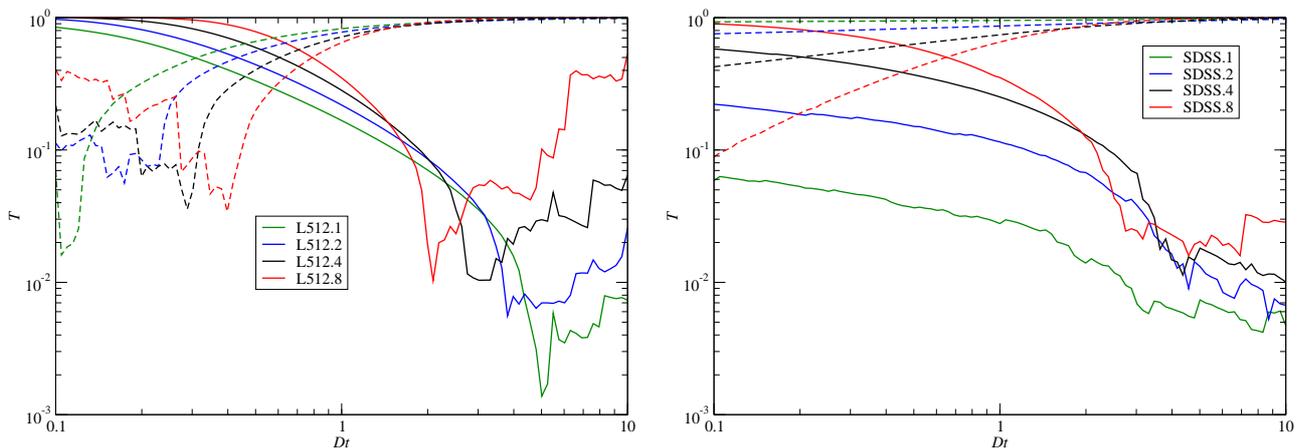

\centering 
\hspace{2mm}  
 \resizebox{0.45\textwidth}{!}{\includegraphics*{L512_fat-D0.eps}}
\hspace{2mm}  
 \resizebox{0.45\textwidth}{!}{\includegraphics*{dr8_fat-D0.eps}}
\\ 
\caption{Fatness factor functions, $\mathcal{T}(D_t)$, of largest
  clusters and voids of the L512 model (left) and of the SDSS galaxy sample
  (right). { Functions for clusters are shown by solid lines, for
    voids by dashed lines.}}
\label{fig:fatness} 
\end{figure*}

\subsection{Fatness factors of clusters and voids}

General geometrical properties of the cosmic web can be studied 
using Minkowski functionals, for pioneering papers see
\citet{Mecke:1994aa}, \citet{Sathyaprakash:1998aa},
\citet{Schmalzing:1999aa}.  Minkowski functionals allow to find their
combination --- shapefinders: thickness, breadth and length of
clusters \citep{Sahni:1998aa, Sheth:2003sp, Shandarin:2004ij}.

In this paper we define a new shape parameter of clusters and voids,
the fatness factor --- the ratio of the volume of clusters/voids to
the maximal possible volume for a given diameter:
\begin{equation}
T = V/D_m^3,
\end{equation}
where
$D_m = \sqrt{\left[(\Delta x)^2 + (\Delta y)^2 + (\Delta
    z)^2\right]/3}$ is the mean diameter of the cluster along
$x,~y,~z$ axes, and $V$ is the volume of the cluster.  A similar
definition is used to calculate fatness factors of voids.  

Fatness factors are dimensionless quantities, and describe the fragile
shape of clusters/voids. Clusters percolate at threshold density
$D_t \approx 5$, and fill only about
$\mathcal{F}_C(P) \approx 10^{-3}$ of space (for smoothing scale
1~\Mpc).  This means that clusters at percolation thresholds are
extended multi-branching low-volume and fragile structures in all
directions.  With the decrease of threshold density the volume of
clusters increases, and their fragility decreases.  Fragile clusters
and voids are illustrated in Figures 14 -- 17 by
\citet{Shandarin:2004ij}.

In Fig.~\ref{fig:fatness2} we present fatness factors of all clusters
and voids of the DM L512.1 sample at percolation thresholds $D_f=5.01$
and $D_t=0.126$, respectively. Fatness factors are shown as functions
of volumes of clusters, $V$, expressed in cubic \Mpc.  A similar
presentation was given by \citet{Shandarin:2006bs} for shapes and
volumes of voids in their analysis of the $\Lambda$CDM model.  Authors
used the term porosity to denote the fragile shape of systems.

We see that fatness factors of largest clusters are
$T_C \approx 10^{-3}$, whereas fatness factors of largest voids are
$T_V \approx 10^{-1}$.  To see the role of smoothing to fatness
factors. the bottom left panel shows fatness factors of the L512.2
sample at percolation threshold $D_t=3.80$, and bottom right panel the
fatness factors of the SDSS.2 samples at the percolation threshold of
the L512.2 sample, $P=3.80$.  Largest clusters of both samples have
fatness factors, $T_C \approx 10^{-2}$.  In all samples the mean
fatness factor decreases with the increase of the volume of
clusters/voids.  Such decrease is larger in cluster samples obtained
with smaller smoothing length, and smaller in void samples.  The
maximal possible value of the fatness factor has a system, filling the
whole possible cubic space for a given diameter, $T=1.0$.  A spherical
system has the fatness factor, $T= \pi/6= 0.524$.
Fig.~\ref{fig:fatness2} shows that smallest clusters of samples L512.2
and SDSS.2 have mean fatness factors around this value.

{ We define fatness factors of largest clusters and voids for
  various threshold levels, $\mathcal{T}(D_t)$, as shape functions of
  the cosmic web.}  Figure~\ref{fig:fatness} shows fatness factor
functions of largest clusters and voids of the L512 model (left), and
of the SDSS galaxy sample (right).  Fatness factors are calculated for
smoothing lengths $R_B=1,~2,~4,~8$~\Mpc.  At high threshold density
largest clusters are relatively small systems, their fatness factors
fluctuate around $10^{-2}$.  Fatness factors have smallest values near
the percolation threshold density.  Below the percolation threshold
density cluster fatness factors start to grow, due to the growth of
cluster filling factors.  Further lowering the threshold density leads
to additional cluster merging, and the volume of the largest cluster
increases continuously.  Thus fatness factors of clusters grow and
reach values close to unity at smallest threshold densities.  Here the
largest cluster fills almost the whole volume of the sample.  In the
threshold density range where clusters/voids are percolated, diameters
of clusters/voids are equal to their maximal length, $D_m = L_0$, thus
in this range $\mathcal{T}(D_t) = \mathcal{F}(D_t)$.

Fatness factor functions of voids of L512 model samples have similar
behaviour, when started from low threshold densities.  At low
threshold densities largest voids are isolated, and their fatness
factors fluctuate around the value $10^{-1}$.  At void percolation
threshold density void volumes start to grow continuously, and fatness
factor function grows towards unity, following the filling factor function.

The behaviour of the fatness factor functions of SDSS clusters is
different from the behaviour of similar functions of model clusters.
At all smoothing lengths fatness factors of SDSS clusters at low
threshold densities are smaller than fatness factors of L512 model
clusters.  This difference is due to the fact that at small threshold
densities model clusters include to their volume low-density filaments
and sheets of dark matter, which are not present as galaxy filaments
in SDSS samples.  Fatness factor functions of SDSS voids are very
different from fatness factor functions of L512 voids.  Voids in SDSS
samples are percolating at all threshold densities, and cover large
volumes.  Thus filling factor and fatness functions of SDSS voids are
always large, especially for small smoothing lengths.

\section{Discussion and summary}

\subsection{Percolation method as a cosmological tool}

The percolation method can be applied in two ways.  A simple
application is to use it as a tool to select certain kind of galaxy
systems from the density field. In this role it was used by
\citet{Einasto:2006kl, Einasto:2007tg}, \citet{Liivamagi:2012aa} and
\citet{Einasto:2017aa} to find superclusters of galaxies for further
more detailed analysis.  Another example is provided by
\citet{Shandarin:2006bs}, who selected voids in a $\Lambda$CDM model
and studied their shapes and sizes.  In this paper we selected
clusters and voids of the density field,  and investigated their 
distributions and fatness properties.

The second possibility is to use the percolation method as a tool to
investigate geometrical properties of the cosmic web.  The extended
percolation analysis as used in this paper is sensitive not only to
the connectivity of high- and low-density regions, but to a number of
other geometrical properties of the web too.  Among these properties
are the presence or absence of faint filaments around high-density
knots, the filamentary character of the web, the deviation of the
density field from the Gaussian one, and the main topological type of
the web.  The extended percolation analysis allows to calculate a
large number of functions, which characterise general geometrical
properties of the density field.  It also allows to define 
important quantitative parameters, as the percolation threshold
density. This parameter depends on the smoothing length using in the
calculation of the density field.

To judge the quality of the percolation method to investigate the
structure of the cosmic web, we have to know how sensitive it is to
various basic parameters of the model, such as the cosmology, given by
the power spectrum of density fluctuations, used to simulate the
cosmic web, and the size of the simulation box.  The $\Lambda$CDM
model is now well established, thus we see no need to vary
cosmological parameters. Instead we calculated percolation functions
for DM simulations in four box sizes from 100 to 1024~\Mpc\ with
identical resolutions $512^3$ particles and cells.  This test showed
that percolation functions of models of different size are very
similar to each other. This stability suggests that properties of the
cosmic web, as found in the present paper, can be applied to the
cosmic web as a whole.  Differences due to the use of several
independent realisations of the model are much smaller than using
different box sizes.

Errors of percolation functions can be estimated on the basis of the
shape of these functions. In this paper we are interested essentially
in general geometrical properties of the cosmic web as an ensemble,
thus exact errors of these functions are of minor importance.  We
calculated errors only for percolation density thresholds, and for
filling factors at mean threshold density, $D_t=1$.  Our analysis has
shown, that these errors are surprisingly small. 

The most significant effect in percolation functions is due to the use
of different smoothing scales.  To understand its influence we used
four values of the smoothing kernel radius.  In this way the effect is
well under control. This sensitivity shows that different samples can be
compared only using identical smoothing scales.  We consider dark
matter as a physical fluid having continuous density distribution.
The fine structure of the dark matter is seen inside DM halos, which
have a characteristic size of 1~\Mpc.  When we investigate the
structure of the cosmic web, then fine details of the web within
galaxy or cluster sized halos and filaments are not important. For
this reason we consider the smoothing length 1~\Mpc\ as the best to
investigate details of the cosmic web.  To get a complex picture of properties of
the cosmic web, the use of different smoothing scales is needed.

At small smoothing scale geometrical properties of clusters and voids
are asymmetrical.  A symmetry of properties of clusters and voids,
observed in several studies \citep{Gott:1986ly}, is valid only when a
large smoothing is applied.

Our study has shown that at low threshold densities, $D_t \le 0.5$,
percolation functions are very sensitive to the presence or absence of
faint filaments between high-density knots.  In this threshold density
interval the fine structure of clusters and voids of model and real
samples are very different. { It is clear that the high sensitivity
of the extended percolation method to the filamentary character of 
faint features of the web can help to investigate the bias between
galaxy and matter distributions.}

At medium threshold densities, $0.5 < D_t < 3$, percolation functions
depend on the true density distribution, as well on observational selection
effects.  Difficulties in the use of flux-limited observational
samples were mentioned by \citet{Martinez:2002fu} in the discussion of
percolation functions.  This could be the reason why percolation
analysis has been made so far mostly for DM models only.  Our
discussion of this effect has shown that the influence of selection
effects to percolation properties is well understood, and we can get
valuable information on clustering properties of the cosmic web by
comparing DM models with observations.

At high threshold densities, $D_t \ge 3$, percolation properties of DM
model clusters are approximately similar to percolation properties of
SDSS cluster samples.  Here one small detail is interesting --- percolation
functions depend not only on general geometry of the density distribution,
but also on the filamentary character of the web, and on the location
of knots in filaments.

\subsection{Summary remarks}

Our work has shown that the extended percolation analysis is a
versatile method to study various geometrical properties of the cosmic
web in a wide range of parameters. We can highlight our findings as
follows. 

\begin{enumerate} 

\item{} Percolation functions of $\Lambda$CDM models of sizes from 100
  to 1024~\Mpc\ are very similar to each other.  This stability
  suggests that properties of the cosmic web, as found in the present
  paper, can be applied to the cosmic web as a whole.

\item{} The percolation threshold of DM models is a function of the
  smoothing length, $R_B$.  The percolation threshold of DM clusters
  is $\log P_C = 0.718 - 0.444 \times \log R_B$, and of DM voids is
  $\log P_V = -0.816 + 0.503 \times \log R_B$, different from
  percolation threshold of random samples, $\log P_0 = 0.00 \pm 0.02$.
 
\item{} Percolation functions depend on the smoothing length to
  calculate density fields. Very small smoothing characterises the
  fine structure inside halos, large smoothing shifts part of matter
  from high-density regions to their surrounding.  

\item{} Percolation functions are  sensitive to very faint
  filaments of the cosmic web, present in DM models, but absent in
  SDSS samples.  At low and medium threshold densities, and at 
  smoothing length $\sim 1$~\Mpc, percolation functions of the SDSS
  sample are  different from percolation functions of DM model
  samples, both for clusters and voids.  The SDSS sample has only one
  large percolating void, which fills almost the whole volume, and
  contains numerous isolated clusters at low threshold densities,
  absent in model samples.  At large threshold densities percolation
  properties of DM and SDSS clusters are similar.

\item{} Percolation analysis allows to calculate fatness of clusters
  and voids: the ratio of the volume of clusters/voids to their
  maximal possible value.  Near percolation threshold the fatness of
  DM clusters is $\approx 10^{-3}$, and of DM voids $\approx 10^{-1}$. 
\end{enumerate}

Differences between distributions of galaxies in the real world and
particles in numerical models were discussed already in early papers
\citep{Joeveer:1978dz, Zeldovich:1982kl, Peebles:2001kl}. The extended
percolation analysis describes these differences in a complex way. 
The application of the extended
percolation method to a more detailed investigation of the formation
of the cosmic web is an interesting goal for future studies.

\begin{acknowledgements} 
 
  We thank Gert H\"utsi, Mirt Gramann, Antti Tamm and Elmo Tempel for
  discussion, { and the anonymous referee for useful suggestions}.
  Our special thank is to Enn Saar who developed the very efficient
  code to find over- and under-density regions.

This work was supported by institutional research funding IUT26-2 and
IUT40-2 of the Estonian Ministry of Education and Research. We
acknowledge the support by the Centre of Excellence ``Dark side of the
Universe'' (TK133) financed by the European Union through the European
Regional Development Fund.  The study has also been supported by
ICRAnet through a professorship for Jaan Einasto.

We thank the SDSS Team for the publicly available data releases.
Funding for the SDSS and SDSS-II has been provided by the Alfred
P. Sloan Foundation, the Participating Institutions, the National
Science Foundation, the U.S. Department of Energy, the National
Aeronautics and Space Administration, the Japanese Monbukagakusho, the
Max Planck Society, and the Higher Education Funding Council for
England. The SDSS Web Site is \texttt{http://www.sdss.org/}.

The SDSS is managed by the Astrophysical Research Consortium for the 
Participating Institutions. The Participating Institutions are the 
American Museum of Natural History, Astrophysical Institute Potsdam, 
University of Basel, University of Cambridge, Case Western Reserve 
University, University of Chicago, Drexel University, Fermilab, the 
Institute for Advanced Study, the Japan Participation Group, Johns 
Hopkins University, the Joint Institute for Nuclear Astrophysics, the 
Kavli Institute for Particle Astrophysics and Cosmology, the Korean 
Scientist Group, the Chinese Academy of Sciences (LAMOST), Los Alamos 
National Laboratory, the Max-Planck-Institute for Astronomy (MPIA), 
the Max-Planck-Institute for Astrophysics (MPA), New Mexico State 
University, Ohio State University, University of Pittsburgh, 
University of Portsmouth, Princeton University, the United States 
Naval Observatory, and the University of Washington.

\end{acknowledgements}

\bibliographystyle{aa} 

\end{document}